\documentclass[aip,jcp,twocolumn,nobibnotes,reprint]{revtex4-1}

\usepackage{graphicx,amsfonts,amsmath,amsbsy,amssymb,hyperref,subfigure}
\newcommand{\refeq}[1]{{Eq.~(\ref{#1})}}
\newcommand{\reffig}[1]{{Fig.~\ref{#1}}}

\DeclareMathOperator{\Tr}{Tr}

\hypersetup{breaklinks,colorlinks=true,linkcolor=blue,citecolor=blue,urlcolor=blue,filecolor=blue}

\begin{document}

\title{Piecewise Interaction Picture Density Matrix Quantum Monte Carlo}

\author{William Van Benschoten}
\author{James~J.~Shepherd}

\email{james-shepherd@uiowa.edu}
\affiliation{Department of Chemistry, University of Iowa}
\date{\today}

\begin{abstract}
The density matrix quantum Monte Carlo (DMQMC) set of methods stochastically
samples the exact $N$-body density matrix for interacting electrons at finite
temperature.  We introduce a simple modification to the interaction picture
DMQMC method (IP-DMQMC) which overcomes the limitation of only sampling one
inverse temperature point at a time, instead allowing for the sampling of a
temperature range within a single calculation thereby reducing the
computational cost.  At the target inverse temperature, instead of ending the
simulation, we incorporate a change of picture away from the interaction
picture.  The resulting equations of motion have piecewise functions and use
the interaction picture in the first phase of a simulation, followed by the
application of the Bloch equation once the target inverse temperature is
reached.  We find that the performance of this method is similar to or better
than the DMQMC and IP-DMQMC algorithms in a variety of molecular test systems.
\end{abstract}

\maketitle

\section{Introduction}

Electrons interacting in the presence of a finite temperature play an important
role in many applications including the study of planetary
cores\cite{militzer_understanding_2016,mazzola_phase_2018}, plasma
physics\cite{mukherjee_hot_2013,zhou_aluminum_2016}, laser
experiments\cite{ernstorfer_formation_2009}, and  condensed phases of
matter\cite{gull_superconductivity_2013,drummond_quantum_2015}.  There has been
a recent push to take methods which are effective for solving ground state
electronic structure problems, especially quantum chemical wavefunction
methods, and adapting them to treat finite temperature. Examples of this
include perturbation
theories\cite{he_finite-temperature_2014,santra_finite-temperature_2017,hirata_finite-temperature_2020,jha_finite-temperature_2020,hirata_finite-temperature_2021}
and coupled cluster
techniques\cite{dzhioev_superoperator_2015,hermes_finite-temperature_2015,hummel_finite_2018,harsha_thermofield_2019,white_time-dependent_2019,shushkov_real-time_2019,white_finite-temperature_2020,peng_conservation_2021,harsha_thermal_2022}.
Other \textit{ab initio} methods under active development include
ft-DFT\cite{karasiev_generalized-gradient-approximation_2012,ellis_accelerating_2021,pittalis_exact_2011,eschrig_t_2010,pribram-jones_thermal_2016}
and various flavors of Green's function
methods\cite{kananenka_efficient_2016,welden_exploring_2016,kas_finite_2017,karrasch_finite-temperature_2010,neuhauser_stochastic_2017,gu_generalized_2020,li_sparse_2020}
such as self-consistent second-order perturbation theory (GF2) and \textit{GW}
theory.

Additionally, embedding theories, which break the calculation up into an
exactly treated subsystem and an approximately treated bath, have been
proposed.\cite{knizia_density_2013,sun_finite-temperature_2020,kretchmer_real-time_2018,tran_using_2019,cui_efficient_2020,zhai_low_2021,tsuchimochi_density_2015,bulik_electron_2014,hermes_multiconfigurational_2019,zgid_finite_2017,lan_generalized_2017,tran_spin-unrestricted_2018,rusakov_self-energy_2019}
There are also a variety of quantum Monte Carlo methods which work with finite
temperature ensembles of electrons, such as path integral Monte
Carlo\cite{dornheim_permutation_2015,militzer_development_2015,larkin_phase_2017,groth_configuration_2017,dornheim_ab_2018,dornheim_fermion_2019,yilmaz_restricted_2020,dornheim_ab_2021},
determinant quantum Monte Carlo
(DQMC)\cite{lee_parallelizing_2012,chang_recent_2015}, finite temperature
auxiliary field quantum Monte Carlo
(ft-AFQMC)\cite{liu_ab_2018,he_finite-temperature_2019,shen_finite_2020,church_real-time_2021,liu_unveiling_2020},
and Krylov-projected quantum Monte Carlo\cite{blunt_krylov-projected_2015}.

The method we use here, density matrix quantum Monte Carlo (DMQMC),
stochastically samples the exact N-body density matrix in a finite
basis.\cite{blunt_density-matrix_2014} It is the finite temperature equivalent
to FCIQMC\cite{booth_fermion_2009}, which has been very successful in treating
ground-state problems to FCI accuracy.  In the original
paper,\cite{blunt_density-matrix_2014} DMQMC calculated thermal quantities for
the Heisenberg model including the energy and Renyi-2 entropy.  Thereafter,
interaction picture DMQMC (IP-DMQMC) was introduced which introduced a change
of picture allowing for the diagonal (and trace) of the density matrix to be
sampled much more accurately than DMQMC.\cite{malone_interaction_2015} It does
so by simulating one temperature at a time and starting at an approximate
density matrix for that temperature.  After being developed to use the
initiator approach,\cite{malone_interaction_2015} which was also adapted from
the ground-state FCIQMC version\cite{cleland_communications_2010}, IP-DMQMC
benchmarked the warm dense electron gas alongside path integral Monte Carlo
approaches to obtain a finite-temperature local density approximation
functional.\cite{dornheim_ab_2017,groth_ab_2017,dornheim_ab_2016} In addition
to these successes, IP-DMQMC showed promise in initial applications to
molecular systems\cite{petras_using_2020} and its sign problem showed to be
similar to that of FCIQMC.\cite{petras_sign_2021} Recently DMQMC has also
inspired a new method, fixed point quantum Monte Carlo, which samples the
ground state density matrix.\cite{chessex_fixed_2022}

In this work we seek to extend IP-DMQMC by continuing the simulation after the
target inverse temperature is reached. We find that continuing to apply the
Bloch equation as the propagator allows for the rest of the
temperature-dependent energy to be found. This is possible because IP-DMQMC
reaches the exact density matrix (on average) once it reaches a target
temperature.  This paper starts with an introduction to DMQMC methods followed
by a derivation of the new piecewise IP-DMQMC propagation equations (which we
call PIP-DMQMC).  Next we test PIP-DMQMC for a set of molecular systems making
comparison with DMQMC, IP-DMQMC, and finite temperature full configuration
interaction (ft-FCI).\cite{kou_finite-temperature_2014} We then explore how
PIP-DMQMC can be combined with the initiator approximation (i-PIP-DMQMC) and
that these can be used to treat larger systems that cannot be exactly
diagonalized. We close by noting how the compute time cost of simulating a
range of evenly-spaced target inverse temperatures in IP-DMQMC scales roughly
as the square of the largest target inverse temperature sampled, while
PIP-DMQMC samples the same range with linear scaling.

\section{Methods}

In this section, we begin with a review of the DMQMC algorithm including key
algorithmic definitions and the initiator approximation.  We then describe how
IP-DMQMC differs from DMQMC and the sum-over-states methods we used here
(ft-FCI and THF). Section \ref{sec:3} introduces our new method PIP-DMQMC. 

\subsection{DMQMC}

The DMQMC set of methods\cite{blunt_density-matrix_2014} was a generalization
of full configuration interaction quantum Monte Carlo
(FCIQMC)\cite{booth_fermion_2009} to solving the $N$-body density matrix at
finite temperature:
\begin{equation}
\hat{\rho}(\beta)=e^{-\beta \hat{H}},
\end{equation}
where $\beta=1/kT$. The density matrix is written in a finite basis in
imaginary time:
\begin{equation}
\hat{\rho}(\beta)\rightarrow \hat{f}(\tau)=\sum_{ij} f_{ij} (\tau)  | D_i \rangle \langle  D_j|,
\end{equation}
where $D_i$ are the orthogonal Slater determinants. These determinants are
formed from the orbitals calculated during a ground state Hartree--Fock (HF)
calculation.  In DMQMC, the density matrix is represented by walkers which are
said to be on sites in the simulation.  One site, labelled with an $i$ and $j$
index, represents each $| D_i \rangle \langle  D_j |$ and the number of walkers
at site $ij$ is proportionate to $f_{ij}$.  The goal is to find
\begin{equation}
E(\beta)=\frac{\Tr[\hat{H}\hat{\rho}(\beta)]}{\Tr[\hat{\rho}(\beta)]},
\end{equation}
by taking an average over $N_\beta$ separate simulations (these are termed
$\beta$ loops).

In DMQMC, the simulation is started from a random distribution of walkers along
the diagonal of the density matrix, which represents the exact density matrix
at high temperature:
\begin{equation}
\hat{f}(\tau=0)=\mathbb{1}.
\end{equation}
The symmetrized Bloch equation is then applied in the form
\begin{equation}
\frac{\mathrm{d} \hat{f}(\tau)}{\mathrm{d} \tau} = -\frac{1}{2}\left[ \hat{H} \hat{f}(\tau)+\hat{f}(\tau) \hat{H} \right],
\label{eq:bloch}
\end{equation}
and using the Bloch equation with a finite $\Delta\tau$ yields the equations of
motion:
\begin{equation}
\begin{split}
f_{ij}&(\tau + \Delta \tau) =
\\
& f_{ij}(\tau)\left[1+\Delta\tau S\right] - \frac{\Delta\tau}{2}\sum_{k}\left[ H_{ik}f_{kj}(\tau)+f_{ik}(\tau)H_{kj}\right],
\end{split}
\label{eq:eom_dmqmc}
\end{equation}
where a constant shift ($S$) is applied to each matrix element, and is
dynamically updated during the simulation to control the walker population.  At
every time step for DMQMC, $f_{ij}(\tau)$ is equivalent to sampling
$\rho_{ij}(\beta)$ at $\beta=\tau$. 

Equation \ref{eq:eom_dmqmc} is interpreted in the DMQMC algorithm by steps
referred to as spawning, cloning/death, and annihilation. This is key for the
computational efficiency of the method and are described in more detail in the
supplementary information.  Spawning stochastically samples the $\sum_k$ taking
advantage of the sparsity of $\hat{H}$ and the death/cloning steps control the
walker population/memory cost. 

The shift is updated at intervals using: 
\begin{equation}
S(\tau + \Delta\tau)=S(\tau)-\frac{\zeta}{A\Delta\tau}\ln\left(\frac{N_w(\tau + \Delta\tau)}{N_w(\tau)}\right).
\label{eq:shift_update}
\end{equation}
In this equation $A$ is the number of imaginary time steps (iterations of
\refeq{eq:eom_dmqmc}) between updates, $N_w$ is the walker population, and
$\zeta$ is a damping parameter. For this study we used $A=10$ and $\zeta=0.05$.

The initiator approximation\cite{cleland_communications_2010} was developed in
FCIQMC and subsequently adapted for DMQMC.\cite{malone_accurate_2016} The
initiator approximation is a way to stabilize the sign problem, (which arises
in numerical methods due to coefficients in the solution having both positive
and negative values).  In DMQMC, two parameters are introduced: a spawned
walker threshhold, $n_\mathrm{add}$, and a excitation number cutoff,
$n_\mathrm{ex}$.  The population of sites occupied in the simulation (the $ij$
indices) is divided into initiators which have a population that is $\ge
n_\mathrm{add}$ \emph{or} an excitation number $\le n_\mathrm{ex}$.  To find
the excitation number for the $ij$ site, the excitations between $D_i$ and
$D_j$ in $| D_i \rangle \langle D_j |$ are used (i.e. if $D_i$ was a double
excitation of $D_j$, this would count as $n_\mathrm{ex}=2$).  Spawning events
to sites without walkers are then allowed only if they come from initiator
sites or two spawning events with the same sign arrive at once to a single
site.  This introduces a population-and-system-dependent systematic error to
the simulation that is removed in the limit of an infinite total walker
population ($N_w\rightarrow\infty$).  This is typically referred to as being
systematically
improvable.\cite{cleland_communications_2010,cleland_study_2011,cleland_taming_2012,shepherd_investigation_2012,booth_towards_2013,thomas_accurate_2015,malone_accurate_2016}
We used $n_{\mathrm{add}}=3.0$ and $n_\mathrm{ex}=2$ in this study, which came
from preliminary investigations and was consistent with previously used values
for the uniform electron gas.\cite{malone_accurate_2016}

\subsection{IP-DMQMC}

In IP-DMQMC the simulation starts with a known density matrix: 
\begin{equation}
\hat{f}(\tau=0)=e^{-\beta_T \hat{H}^{(0)}},
\label{eq:init_ip}
\end{equation}
where $\hat{H}^{(0)}$ is the diagonal of $\hat{H}$.  Then, the simulation
proceeds in imaginary time aiming to sample the exact density matrix at a
certain $\beta$ (the target $\beta$ or $\beta_{T}$).  Over the simulation
IP-DMQMC samples
\begin{equation}
\hat{f}(\tau)=e^{-\left(\beta_T-\tau\right)\hat{H}^{(0)}}e^{-\tau \hat{H}}
\label{eq:fIP}
\end{equation}
instead of sampling $e^{-\tau \hat{H}}$.  Each IP-DMQMC calculation requires a
$\beta_{T}$ to be specified which is the inverse temperature at which we wish
to calculate the density matrix.

The initial density matrix $e^{-\beta \hat{H}^{(0)}}$ is stochastically
sampled.  The protocol is as follows. First, select a determinant by occupying
$N$ orbitals at random, and then accepting this selection according to the
product of its normalized Fermi-Dirac weights:
\begin{equation}
P=\prod_{p \in \mathrm{occ}}n_p \prod_{p \in \mathrm{unocc}}1-n_p,
\label{eq:FD_prob}
\end{equation}
where $p$ refers to an orbital index. The abbreviations $\mathrm{occ}$ and
$\mathrm{unocc}$ respectively refer to occupied and unoccupied orbital indices.
The normalized Fermi-Dirac weights are calculated as:
\begin{equation}
n_i = \frac{1}{e^{\beta\left(\epsilon_i-\mu\right)}+1}
\label{eq:FD},
\end{equation}
where $\epsilon_{i}$ are the single--particle energies (ground--state HF
orbital eigenvalues), and $\mu$ is a temperature-dependent chemical potential
calculated so that the weights sum to the particle
number~\cite{malone_interaction_2015}.  Accepting only those orbital
configurations which conserve the particle number and symmetry ensures a
canonical distribution is obtained. 

The Fermi-Dirac weights sample $e^{-\beta \hat{H}^\prime}$, where
$\hat{H}^\prime$ is diagonal in the Slater Determinant basis and is made from
non-interacting orbitals that have as their orbital energies the HF
one-particle eigenvalues.  Noting that $e^{-\beta \hat{H}^\prime}\ne e^{-\beta
\hat{H}^{(0)}}$, the walker population on that site is then set to the
difference between the $e^{-\beta \hat{H}^{(0)}}$ and the $e^{-\beta
\hat{H}^\prime}$, normalized such that selection of $H^{(0)}_{00}$ is given a
weight of 1, then the population is stochastically rounded to a predefined
cutoff (0.01 in this work).\cite{GCI} Annihilation is performed on the spawned
walkers, prior to propagation, once the desired walker population is reached
from initialization.

The propagator in IP-DMQMC is
\begin{equation}
\frac{\mathrm{d} \hat{f}(\tau)}{\mathrm{d}\tau} = \hat{H}^{(0)}\hat{f}(\tau)-\hat{f}(\tau)\hat{H},
\label{eq:3}
\end{equation}
and because the matrix \refeq{eq:fIP} contains $\hat{H}^{(0)}$, $\hat{H}^{(0)}$
appears in our propagator.  The absence of a $\frac{1}{2}$ when compared to
DMQMC (\refeq{eq:bloch}) is because the propagator is asymmetric to save the
computational cost of storing additional factors~\cite{malone_quantum_2017}.
Using the propagator with a finite $\Delta\tau$, and including a constant shift
($S$) for controlling the population the equations of motion is written as:
\begin{equation}
\begin{split}
f_{ij}&(\tau + \Delta \tau) =
\\
&f_{ij}(\tau)\left[1+\Delta\tau S\right] -\Delta\tau\sum_{k}\left[-H_{ik}^{(0)}f_{kj}(\tau)+f_{ik}(\tau)H_{kj}\right].
\end{split}
\label{eq:eom_ipdmqmc}
\end{equation}
In IP-DMQMC, $f_{ij}(\tau)$ is equivalent to sampling $\rho_{ij}(\beta)$ only
at $\tau=\beta_T$.

\subsection{ft-FCI and THF}

In this work, it is useful to have exact finite temperature electronic energies
to benchmark the QMC methods.  Therefore, we use a sum-over-states method to
generate the exact energy (within a basis), known as
ft-FCI.\cite{kou_finite-temperature_2014} The finite temperature energy is
\begin{equation}
E=\frac{\sum_i E_i e^{-\beta E_i}}{\sum_i e^{-\beta E_i}},
\label{eq:ftfci}
\end{equation}
where $E_i$ are the energy eigenstates resulting from the exact diagonalization
(FCI) of $\hat{H}$.

In addition to ft-FCI, we calculate an energy referred to as thermal HF (THF),
which is a sum over Boltzmann weighted Slater determinants: 
\begin{equation}
E^{(0)}=\frac{\sum_i E_i^{(0)} e^{-\beta E_i^{(0)}}}{\sum_i e^{-\beta E_i^{(0)}}},
\label{eq:THF}
\end{equation}
$E_i^{(0)}$ is the energy of a single Slater determinant comprised of the
canonical orbitals from a ground-state HF calculation.  In our case, this means
$E_i^{(0)}=H_{ii}$, or the diagonal of the FCI Hamiltonian prior to
diagonalization.  To calculate the thermal correlation energy, we subtract THF
from ft-FCI. When the correlation energy is negative, ft-FCI is lower in energy
than THF.

\section{PIP-DMQMC}
\label{sec:3}

The benefit of IP-DMQMC is that the starting density matrix is much closer to
the density matrix at $\beta_T$ and the change of the density matrix over the
simulation is reduced.  However, the limitation of this approach is that a
single target temperature must be simulated at a time and, consequently, this
means that more computer time is spent in the calculation trying to reach
$\beta_T$ than collecting statistics at $\beta_T$.  Our method here is designed
to overcome this limitation. 

Noting that the density matrix in an IP-DMQMC simulation is exact at
$\tau=\beta_T$, \emph{i.e.}
\begin{equation}
f_{ij}(\tau=\beta_T)=\rho_{ij}(\beta_T),
\end{equation}
application of the Bloch equation on this IP-DMQMC density matrix would allow
us to determine $\hat{\rho}(\beta_T+\Delta \tau)$.  Repeated application of the
Bloch equation will generally allow the density matrix at $\beta>\beta_T$ to be
found.

In piecewise IP-DMQMC (PIP-DMQMC) we therefore start with the same density
matrix as in IP-DMQMC, 
\begin{equation}
\hat{f}(\tau=0)=e^{-\beta_T \hat{H}^{(0)}},
\label{eq:THF2}
\end{equation}
and the propagator can be expressed as,
\begin{align}
\frac{\mathrm{d} \hat{f}(\tau)}{\mathrm{d}\tau}=&\Theta(\beta_{T}-\tau)\left[\hat{H}^{(0)}f(\tau)-\hat{f}(\tau)\hat{H}\right]\nonumber
\\&-\frac{1}{2}\left[1-\Theta(\beta_{T}-\tau)\right]\left[\hat{H}\hat{f}(\tau)+\hat{f}(\tau)\hat{H}\right],
\label{eq:pip_heaviside}
\end{align}
where $\Theta(x)$ is a Heaviside step function defined as being equal to 1 for
$x > 0$ and 0 otherwise.  We note that the propagator is always asymmetric in
the IP-DMQMC part of the simulation as symmetrizing IP-DMQMC is
non-trivial.\cite{malone_quantum_2017} In the Bloch equation above, the
propagator is explicitly symmetrized, which is where the factor of
$\frac{1}{2}$ comes from.\newpage

A formulation where the Bloch equation is asymmetric is also possible: 
\begin{align}
\frac{\mathrm{d} \hat{f}(\tau)}{\mathrm{d}\tau}=&\Theta(\beta_{T}-\tau)\left[\hat{H}^{(0)}f(\tau)-\hat{f}(\tau)\hat{H}\right]\nonumber
\\&-\left[1-\Theta(\beta_{T}-\tau)\right]\hat{f}(\tau)\hat{H}.
\label{eq:asympip_heaviside}
\end{align}
Using these propagators with a finite $\Delta\tau$, the equations of motion are:
\begin{widetext}
\begin{equation}
f_{ij}(\tau+\Delta\tau)=
\begin{cases}
f_{ij}(\tau)[1+\Delta\tau S] - \Delta \tau \sum_{k}\left[-H^{(0)}_{ik}f_{kj}(\tau)+f_{ik}(\tau)H_{kj}\right] & \tau < \beta_T
\\
f_{ij}(\tau)[1+\Delta\tau S] - \frac{\Delta \tau}{2}\sum_{k}\left[H_{ik}f_{kj}(\tau)+f_{ik}(\tau)H_{kj}\right] & \tau \geq \beta_T
\end{cases};
\label{eq:symPIP}
\end{equation}
\begin{equation}
f_{ij}(\tau+\Delta\tau)=
\begin{cases}
f_{ij}(\tau)[1+\Delta\tau S] - \Delta \tau \sum_{k}\left[-H^{(0)}_{ik}f_{kj}(\tau) +f_{ik}(\tau)H_{kj}\right] & \tau < \beta_T
\\
f_{ij}(\tau)[1+\Delta\tau S] - \Delta \tau\sum_{k} f_{ik}(\tau)H_{kj} & \tau \geq \beta_T
\end{cases}.
\label{eq:asymPIP}
\end{equation}
\end{widetext}
In \refeq{eq:symPIP}, the IP-DMQMC equations of motion (\refeq{eq:eom_ipdmqmc})
is used for the sub-domain before the target beta is reached ($\tau<\beta_{T}$)
and the symmetric Bloch equation (\refeq{eq:eom_dmqmc}) is used subsequently
($\tau \ge \beta_{T}$).  In \refeq{eq:asymPIP}, the formalism drops the
symmetrization in the Bloch equation.

Which of these equations of motion, symmetric or asymmetric, gets used tends to
be based on preliminary calculations and prior knowledge.  As in the original
DMQMC algorithm, symmetric propagation should generally result in less
stochastic noise but may also raise the plateau.\cite{petras_sign_2021} This is
consistent with what we found in preliminary calculations: the asymmetric
propagation is effective without the initiator approximation and symmetric
propagation is effective with the initiator approximation. 

In PIP-DMQMC, $f_{ij}(\tau)$ is equivalent to sampling $\rho_{ij}(\beta)$ at
$\tau\ge\beta_T$.  The key result here is that piecewise propagation has the
potential benefits of IP-DMQMC -- skipping initialization on the identity
matrix -- while also allowing for continued propagation using the Bloch
equation to higher values of $\beta$.  Below, we mainly test PIP-DMQMC
simulations using $\beta_T=1.0$, but we also examine the effects of changing
$\beta_T$.  Finally, we also note in passing that when the total $\beta$
simulated in PIP-DMQMC is $\beta_T$, the simulation reverts back to IP-DMQMC,
thus we would expect these to be equivalent when the random number seed is
fixed. 

\section{Calculation Details}

The PIP-DMQMC algorithm was implemented in the
HANDE-QMC\cite{spencer_hande-qmc_2019} package.  The molecular systems used
were: Be/aug-cc-pVDZ, BeH$_2$ Be/cc-pVDZ H/DZ, equilibrium H$_8$/STO-3G,
equilibrium H$_4$/cc-pVDZ, stretched H$_8$/STO-3G, N$_2$/STO-3G, LiF/STO-3G,
CO/STO-3G, HCN/STO-3G, HBCH$_2$/STO-3G, H$_2$O/cc-pVDZ and CH$_4$/cc-pVDZ.
Geometries are provided in the supplementary information and come from a
variety of sources.\cite{bernath_vibration-rotation_2002, petras_sign_2021,
booth_fermion_2009, herzberg_electronic_1966, frisch_gaussian_2009,
lovas_diatomic_2002, wharton_microwave_1963, johnson_nist_2006} Molecular
integral files for the one- and two-particle interactions organized by orbital
indices are generated with Molpro.\cite{werner_molpro_2019} HANDE-QMC was used to run
DMQMC, IP-DMQMC and PIP-DMQMC and to generate the FCI and $\hat{H}^{(0)}$
eigenspectrum for ft-FCI and THF respectively. A python script was used to
perform the sum-over-states method for ft-FCI and THF.  Other simulation
parameters are included in our data repository and supporting information.

\section{Results and Discussion}

In order to validate PIP-DMQMC, we set a target accuracy for the energy of a 1
millihartree difference to the ft-FCI energy, which is the standard for high
accuracy approaches in QMC for the ground state. We note in passing that this
may be too high of an accuracy for finite temperature applications. We measured
the accuracy for a range of test systems which we could treat with
exact diagonalization. The number of density matrix elements (the square of the
determinants in the space) for these systems range from $10^5$ to $10^8$. As
well as PIP-DMQMC, four other calculations were performed: DMQMC, IP-DMQMC,
ft-FCI, and THF. For THF, the orbitals and eigenvalues are frozen in the ground
state. This is opposed to the traditional thermal Hartree--Fock method where the
orbitals and eigenvalues are self consistently calculated at the given
temperature. For this reason, THF is a thermal Hartree--Fock like mean-field
method.\cite{malone_accurate_2016}

In each of the DMQMC-type calculations, the walker population was set above the
DMQMC plateau, \cite{petras_sign_2021} which is a system-specific number of
walkers required to overcome the sign problem, and the initiator adaptation was
not used.  The walker numbers used ranged from $5\times10^5$ to $10^7$
depending on the plateau estimated for each system by running a single
calculation with no shift ($S=0$ in \refeq{eq:eom_dmqmc} throughout the
simulation). Energies are reported as averages over 100 $\beta$-loops
($N_\beta=100$).

Preliminary investigations showed that systematic errors were slightly smaller
in magnitude for asymmetric propagation (\refeq{eq:asymPIP}) in PIP-DMQMC (with
no initiator approximation) compared to symmetric propagation
(\refeq{eq:symPIP}). This is consistent with the asymmetric
DMQMC propagators having a smaller plateau, and for this reason we used
asymmetric propagation (\refeq{eq:asymPIP}) in these
tests.\cite{petras_sign_2021}

\begin{figure}
\includegraphics[width=0.4\textwidth, height=\textheight, keepaspectratio]{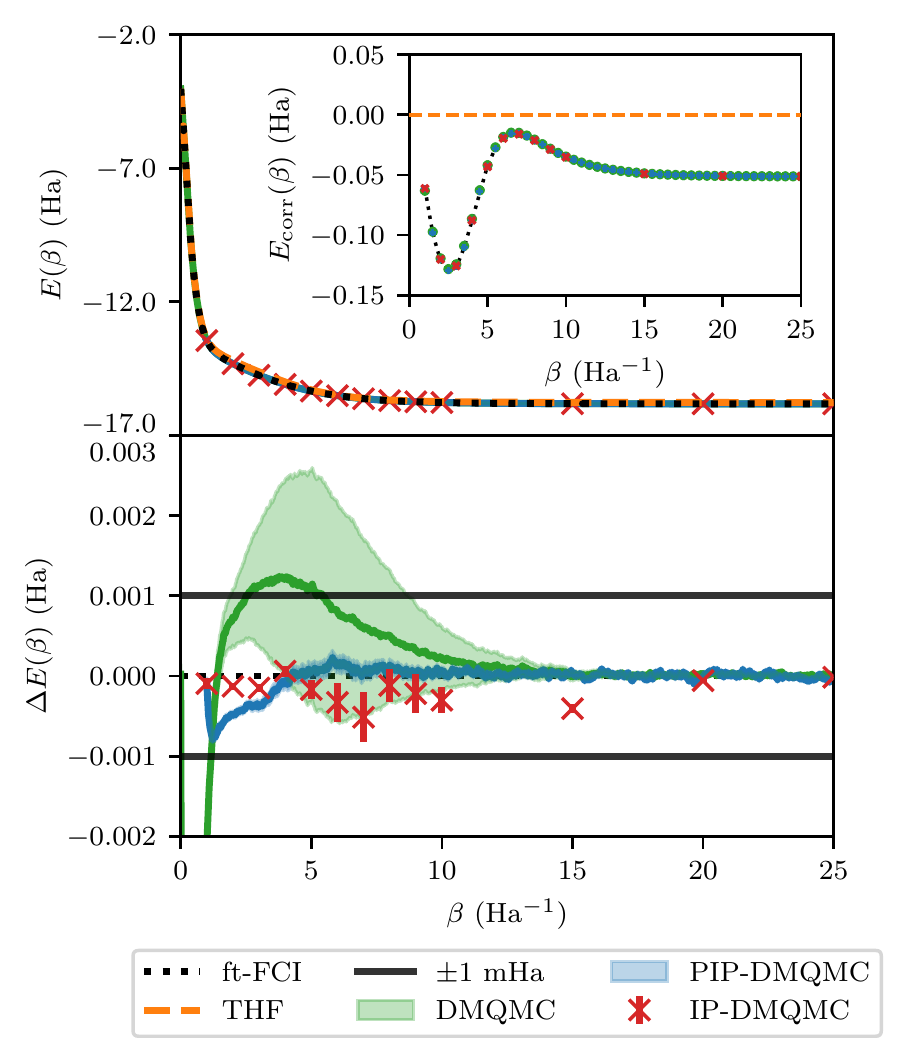}
\caption{Finite temperature energies for BeH$_2$ Be/cc-pVDZ H/DZ from a variety
of methods. In the top panel, total energies are shown plotted against $\beta$.
Energies in the inset are calculated by taking the difference to THF, which is
found with \refeq{eq:THF}. The inset DMQMC and PIP-DMQMC
data are re-sampled every 50 points starting from $\beta=1$. In the bottom
panel, energy differences with respect to ft-FCI are shown (when $\Delta E$ is
positive, ft-FCI is lower in energy). Data here show asymmetric PIP-DMQMC
propagation (\refeq{eq:asymPIP}). The PIP-DMQMC simulations were initialized
with $\beta_{T}=1.0$, allowing data collection at $\beta \geq 1.0$.} 
\label{fig:1}
\end{figure}

A representative example of our set, the BeH$_2$ molecule, is analyzed in
\reffig{fig:1}.  In \reffig{fig:1} (top panel), PIP-DMQMC is found to agree
with DMQMC, IP-DMQMC, and ft-FCI by visual inspection. This was true across the
whole data set but only corresponds to an accuracy of $\sim 0.2$ Ha because the
scale is so large.  In \reffig{fig:1} (bottom panel), we take a closer look at
the systematic energy differences by subtracting the ft-FCI energy.  For the
range of $\beta$ values studied, PIP-DMQMC consistently achieves equivalent or
improved accuracy compared to DMQMC and IP-DMQMC, and its systematic error lies
within 1mHa.  In particular, in common with IP-DMQMC, it is able to improve
upon the `shouldering' seen in this DMQMC line -- an increase in stochastic
error at intermediate $\beta$ values -- which comes from loss of information
from the diagonal of the density in DMQMC (this is which was what prompted the
development of IP-DMQMC)~\cite{malone_interaction_2015}.  PIP-DMQMC generally
performs comparably with IP-DMQMC with the error for both methods falling well
within the stated 1mHa accuracy target, though there is a difference in
performance at $\beta$ values between 5 and 10.

\begin{figure}
\includegraphics[width=0.49\textwidth, height=\textheight, keepaspectratio]{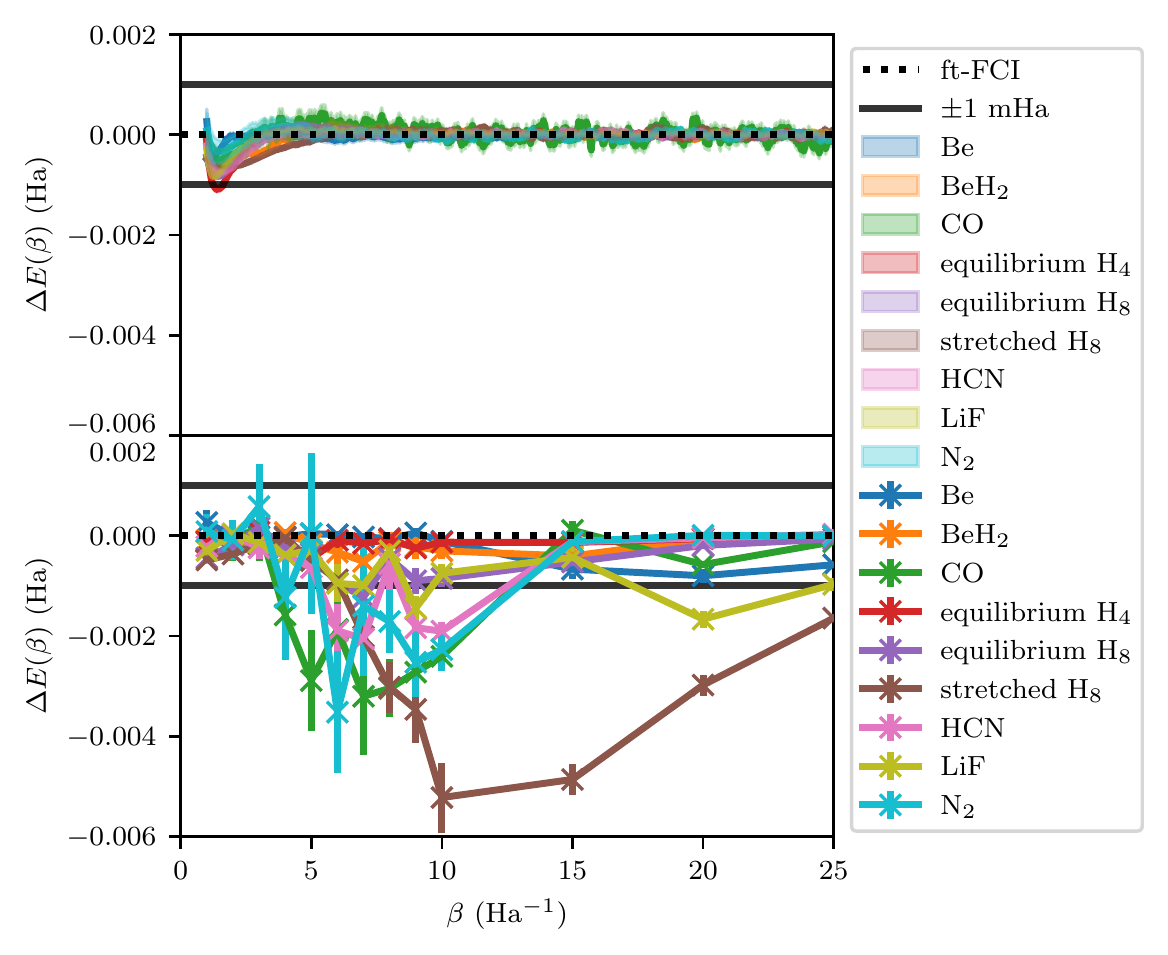}
\caption{Finite temperature energy differences to ft-FCI are shown for a
variety of test systems for PIP-DMQMC (in the top panel) and IP-DMQMC (in the
bottom panel). Data here show asymmetric PIP-DMQMC propagation
(\refeq{eq:asymPIP}). The PIP-DMQMC simulations were initialized with
$\beta_{T}=1.0$, allowing data collection at $\beta \geq 1.0$.} 
\label{fig:2}
\end{figure}

Figure \ref{fig:2} compares IP-DMQMC and PIP-DMQMC by looking at all of the
test set data.  In each case, differences were found between the method and
ft-FCI and we are paying particular attention to the mHa error threshhold.
Here, we do find that IP-DMQMC has a drift towards lower values at intermediate
$\beta$ regimes in most cases which recovers at large $\beta$. A prominent
example of this visible in \reffig{fig:2} is stretched H$_8$, with a maximum
deviation at around $\beta=10$.  Similar to DMQMC, this appears to be related
to how the simulation is initialized; if a particular state is unlikely to be
chosen there can be errors due to under-sampling. This would ultimately be
remedied if enough $\beta$ loops were run.  PIP-DMQMC appears to mostly remedy
this, though we note that PIP-DMQMC does tend to have a dip in energy near its
crossover point (where it switches propagator). Examples of this are BeH$_2$ (\reffig{fig:1}), and CO in the supporting information.

Overall, therefore, we can conclude that PIP-DMQMC achieves just as good if not
better energies than DMQMC and IP-DMQMC across our test set.

\begin{figure}
\begin{center}
\subfigure[\mbox{}]{\includegraphics[width=0.48\textwidth, height=\textheight, keepaspectratio]{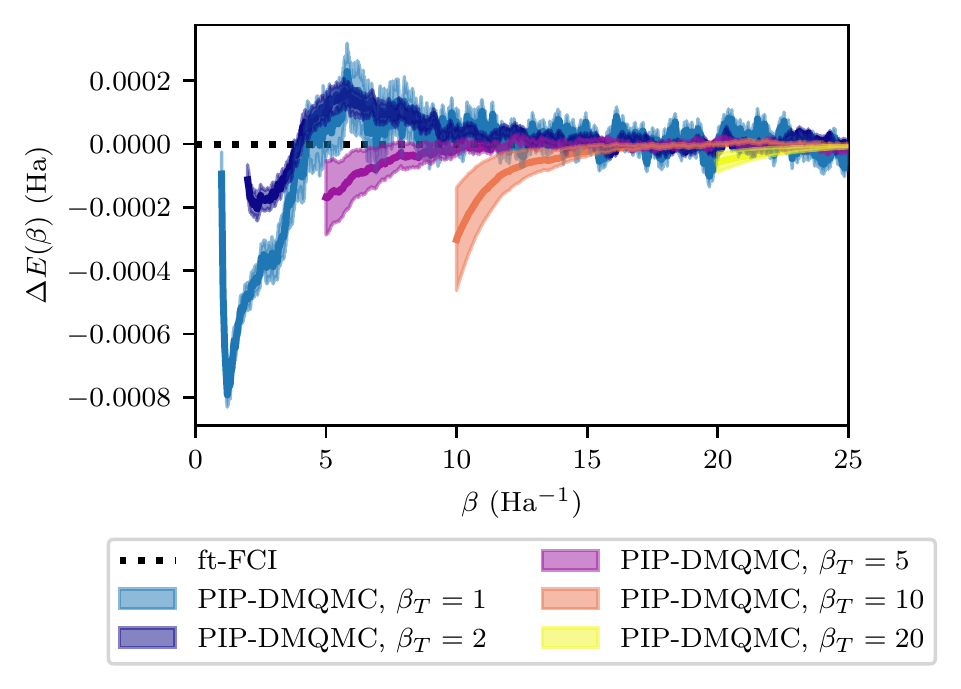}
\label{fig:betaT_scanA}}
\subfigure[\mbox{}]{\includegraphics[width=0.48\textwidth, height=\textheight, keepaspectratio]{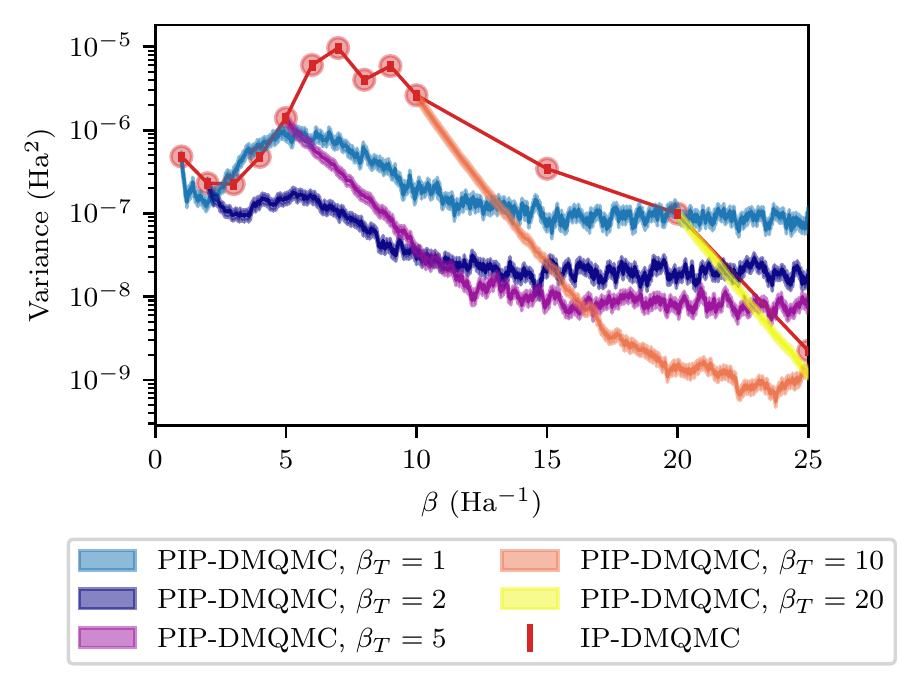}
\label{fig:betaT_scanB}}
\caption{Energy accuracy and variance of the PIP-DMQMC method using several
$\beta_{T}$ values (given in the legends) for BeH$_2$ Be/cc-pVDZ H/DZ. The
impact of changing $\beta_T$ for PIP-DMQMC is evaluated using: (a) the energy
difference to the ft-FCI energy, and (b) the variance of the energy estimate
compared to the IP-DMQMC variance. The $\beta_{T}$ used for IP-DMQMC are the
same as the $\beta$ shown in the figure. PIP-DMQMC and IP-DMQMC have the same
variance when the two share $\beta_{T}$. For $\beta>\beta_{T}$, the PIP-DMQMC
variance is generally below that of IP-DMQMC. The PIP-DMQMC data shown are from
asymmetric propagation (\refeq{eq:asymPIP}). All data were collected using
approximately ten million walkers ($N_w=10^7$) and averaged over 100 $\beta$
loops ($N_\beta=100$).}
\label{fig:3}
\end{center}
\end{figure}

For all data presented so far, we have used a $\beta_{T}=1.0$ for PIP-DMQMC.
Therefore it is worthwhile checking our conclusions for a range of $\beta_{T}$
values to ensure they are not dependent on $\beta_{T}$.  In
\reffig{fig:betaT_scanA} we plot PIP-DMQMC data for a range of $\beta_{T}$
values from $\beta_{T}=1.0$ to $\beta_{T}=20.0$ using \refeq{eq:asymPIP} for
the BeH$_2$ system.

The accuracy of PIP-DMQMC tended to remain similar or improve across the
$\beta_{T}$ used, supporting the prior conclusions made about the method.  For
the energy difference at $\beta=\beta_{T}$, the accuracy tended to be similar
across all the $\beta_{T}$ simulated. For the energy difference at
$\beta>\beta_{T}$, the accuracy in general remained similar or shows slight
improvement. The one exception being for $\beta_{T}=1.0$ where a notable
decrease in accuracy is initially observed. Crucially all lines stay within the
target accuracy of $\pm 1$ mHa, which is too large to be shown on the plot
range used in \reffig{fig:betaT_scanA}.

In addition to investigating the accuracy of PIP-DMQMC using several
$\beta_{T}$, we investigate the variance of PIP-DMQMC in
\reffig{fig:betaT_scanB}.  Both PIP-DMQMC and IP-DMQMC methods have the same
variance at $\beta=\beta_{T}$ because we are using the same random number seed.
In general the variance peaks at intermediate $\beta$. Comparing the variance
from PIP-DMQMC (run beyond the target $\beta$) to the variance from IP-DMQMC,
we see, in general, that the PIP-DMQMC variance is similar to or smaller than
IP-DMQMC.  This is to say that PIP-DMQMC's use of the Bloch equation tends to
decrease variance relative to IP-DMQMC at $\beta>\beta_{T}$.  One reason for
this could be the way in which IP-DMQMC is initialized.  We first recall that
the Fermi-Dirac weights are used to select the diagonal matrix elements and
that then these are re-weighted for the difference between the Fermi-Dirac
density matrix and the Hartree--Fock density matrix (\refeq{eq:THF2}). The
re-weighting happens through spawning a certain number of walkers on this
element.  When the Fermi-Dirac density matrix is a poor match for the
Hartree--Fock density matrix, the re-weighting will be large.  However this, in
turn, limits the high-energy rows that can be included in the simulation.  Over
successive $\beta$ loops, our data are consistent with the idea that the
variance due to this initialization is larger than propagating from an earlier
$\beta$ using the Bloch equation.  The consequence of this is that more $\beta$
loops may be required to convergence the stochastic error in IP-DMQMC as
compared with PIP-DMQMC.

\begin{figure}
\includegraphics[width=0.4\textwidth, height=\textheight, keepaspectratio]{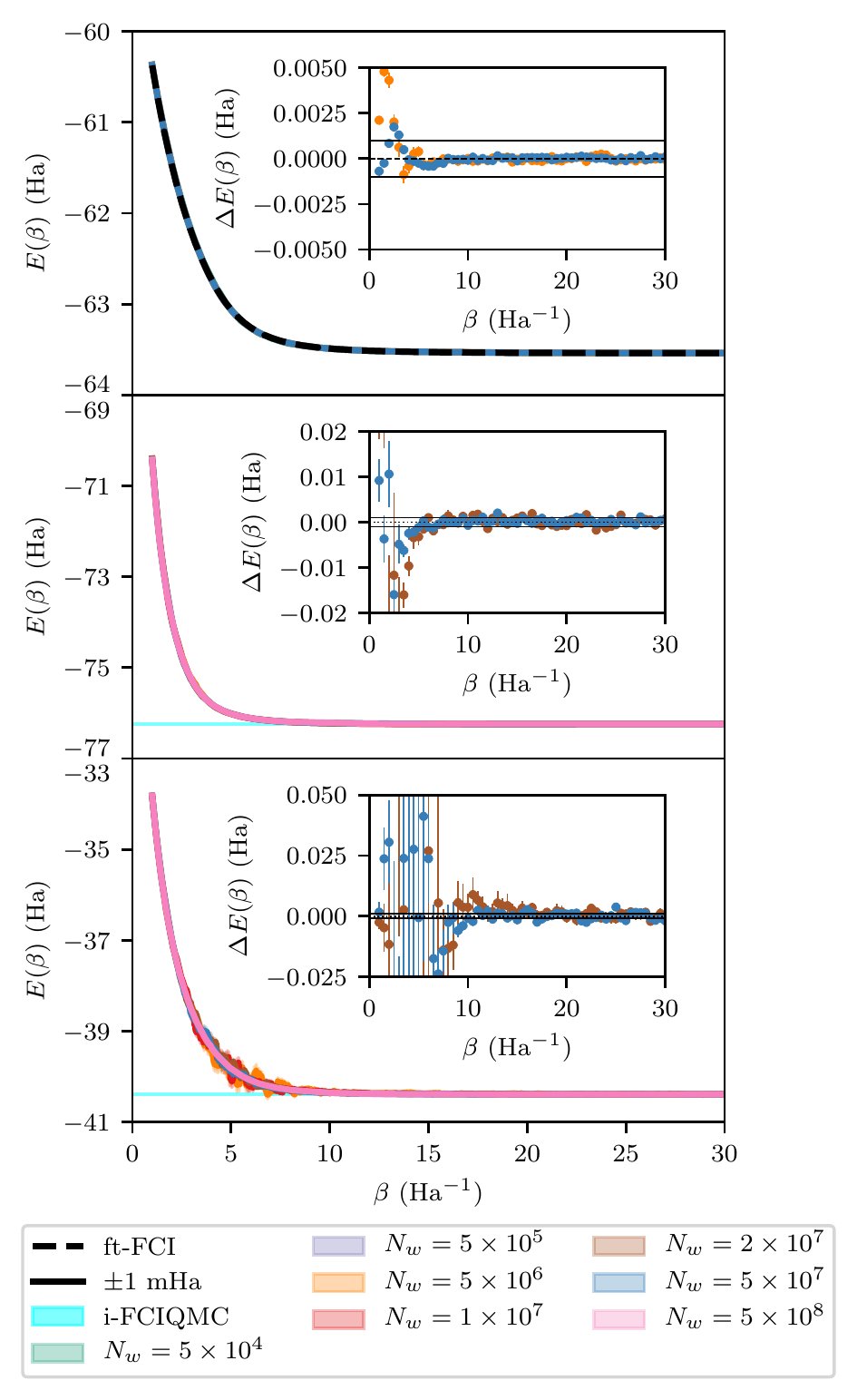}
\caption{Graphs to show initiator convergence for symmetric i-PIP-DMQMC
(\refeq{eq:symPIP}) with increasing walker populations for HBCH$_2$/STO-3G (top
panel), H$_2$O/cc-pVDZ (middle panel) and CH$_4$/cc-pVDZ (bottom panel).  Not
all walker populations were run for all systems.  Except for methane, the
walker numbers fully overlap on the main graph.  To show initiator convergence,
the inset shows the energy calculated as a difference to the simulation with
the largest walker population for CH$_4$ and H$_2$O, and ft-FCI for HBCH$_2$.
The inset only shows the two largest walker populations to make trends more
visible. Inset data are re-sampled every 50 points starting from $\beta=1$.
The largest walker population was $5\times 10^8$ for H$_2$O and CH$_4$.
$N_\beta$ was adjusted between population to give comparable error bars, the
details of which are found in the supplementary information.  The
$N_w=5\times10^7$ for HBCH$_2$ and $N_w=5\times10^8$ for H$_2$O contains one
less $\beta$-loop compared to the other systems matching $N_w$ data sets. The
PIP-DMQMC simulations were initialized with $\beta_{T}=1.0$, allowing data
collection at $\beta \geq 1.0$.}
\label{fig:4}
\end{figure}

Our next test is to find out whether the initiator approximation works well
with PIP-DMQMC. The initiator approximation is important for treating larger
systems because the critical walker population (also known as a plateau height)
rapidly grows beyond what we can store. We found that adding to the initiator
space when the site has a population of 3 or greater ($n_\mathrm{add}=3$) and
having states be initiators if the bra and ket only differ by a double
excitation ($n_\mathrm{ex} = 2$) gives reasonable results; these values were
consistent with previous calculations in the uniform electron
gas.\cite{malone_accurate_2016} As $n_\mathrm{ex}$ causes the simulation to
have a plateau again, we also must make sure we are above the plateau to
overcome the sign problem.

In \reffig{fig:4}, we show the results of the initiator
adaptation with PIP-DMQMC (i-PIP-DMQMC) on three systems: HBCH$_2$, H$_2$O, and
CH$_4$. These calculations were run with a variety of walker numbers so that
we could check for initiator convergence, which occurs when the walker number
is increased without the energy changing.\cite{cleland_communications_2010,
booth_breaking_2011, shepherd_investigation_2012} In preliminary calculations,
we noted relatively little difference between the two different modes of
propagation and decided to use symmetric propagation (\refeq{eq:symPIP}) as
this had been shown to reduce stochastic noise in previous
studies.\cite{malone_quantum_2017} 

In general, we see that the plots for different systems at different walker
numbers generally overlay one another.  The exception to this is low walker
populations for CH$_4$, and this is an example where the population is below
the effective $n_\mathrm{ex} = 2$ plateau height at intermediate $\beta$.
Each plot includes an inset where the
difference is taken to the largest walker population for H$_2$O and CH$_4$ and
to ft-FCI for HBCH$_2$.  For HBCH$_2$, we show the convergence to ft-FCI in the
inset. At the highest walker population, the systematic error generally falls
to $\sim1$mHa for most $\beta$ values.  At its peak, the systematic error
reaches $\sim2$mHa at $\beta$ at the highest population.  For H$_2$O and
CH$_4$, where we do not have ft-FCI results available, the inset instead shows
the energy difference to the largest walker number (which is taken to be the
best estimate we have).  We find that, for high $\beta$, the initiator error is
well converged.  Between $N_w=5\times10^7$ and $N_w=5\times10^8$, the energy
difference is submillihartree above $\beta>5$ and $\beta>10$ for H$_2$O and
CH$_4$ respectively.  For the smaller $\beta$ values for both systems, the
stochastic error rises sharply to an order of magnitude higher (i.e. $\sim
0.01$Ha).  Any systematic initiator error is difficult to estimate due to this
increase in stochastic error.  In general, this shows significant promise for
PIP-DMQMC and highlights the advantage of having so much data over the $\beta$
range available. 

One final consideration is that of computational cost. The cost ($C$) of a
simulation can be considered to be proportionate to the total amount of $\beta$
that is simulated. In IP-DMQMC, sampling from an initial inverse temperature 0
to a final inverse temperature $\beta$ with even spacing ($\Delta\beta$)
results in a cost: $C \propto \beta \left(\frac{\beta}{2 \Delta \beta}
+\frac{1}{2}\right)$. By contrast, PIP-DMQMC has a scaling of: $C\propto\beta$,
which is the same as IP-DMQMC when $\Delta\beta=1$ and a temperature range is
not being sampled. IP-DMQMC increases in cost as more $\beta$ values are
sampled and has an asymptotic scaling of $\mathcal{O}(\beta^2)$, while
PIP-DMQMC costs $\mathcal{O}(\beta)$. In this way, PIP-DMQMC can be seen to
both have an improved cost and resolution of the energy as a function of
$\beta$.

\section{Conclusions}

In summary, we have introduced a piecewise generalization of the interaction
picture propagator in density matrix quantum Monte Carlo, which leads to us
being able to sample a wide range of temperatures with a single calculation.
In proof of concept calculations on a variety of molecular systems, PIP-DMQMC
is generally at least as accurate as IP-DMQMC and DMQMC.  The result is a
reduction in calculation cost to sample energies at 0 to $\beta$ from
$\mathcal{O} (\beta^2)$ to $\mathcal{O} (\beta)$ while also obtaining more data
points.  Furthermore, we see an improvement of sampling statistics especially
at large $\beta$ and have the ability to see more convergence data for the
initiator approximation.  We hope that others can use the water and methane
calculations for benchmarking their calculations. 

Taken together, our results show that PIP-DMQMC is a promising development for
DMQMC.  Our long-term goal is to make DMQMC a resource for benchmarking finite
temperature methods development \cite{he_finite-temperature_2014,
santra_finite-temperature_2017, hirata_finite-temperature_2020,
jha_finite-temperature_2020, hirata_finite-temperature_2021,
dzhioev_superoperator_2015, hermes_finite-temperature_2015, hummel_finite_2018,
harsha_thermofield_2019, white_time-dependent_2019, shushkov_real-time_2019,
white_finite-temperature_2020, peng_conservation_2021, harsha_thermal_2022} as
well as describing electronic structure phenomena at finite temperature.  With
this in mind, two limitations of our study were that we did not systematically
study convergence with initiator parameters or with starting initialization.
These are interrelated because they both cause systematic biases in the energy
and so efforts have to be made to separate out and measure the two effects. We
are in the process of preparing a forthcoming manuscript on this
topic.\cite{vanbenschoten_row_2022}

\section{Acknowledgements}

Research was supported by the U.S. Department of Energy, Office of Science, Office of Basic Energy Sciences Early Career Research Program (ECRP) under Award Number DE-SC0021317. 

This research also used resources from the University of Iowa and the resources of the National Energy Research Scientific Computing Center, a DOE Office of Science User Facility supported by the Office of Science of the U.S. Department of Energy under Contract No. DE-AC02-05CH11231 (computer time for calculations only).

This research used a locally modified version of the development branch of HANDE-QMC.\cite{spencer_hande-qmc_2019}
The code will be queued for public release after the manuscript is published. 
For the purposes of providing information about the calculations used, files will be deposited with Iowa Research Online (IRO) with a reference number [to be inserted at production].
 
\section{Data Availability}
The data that supports the findings of this study are available within the article.


\begin{thebibliography}{91}%
\makeatletter
\providecommand \@ifxundefined [1]{%
 \@ifx{#1\undefined}
}%
\providecommand \@ifnum [1]{%
 \ifnum #1\expandafter \@firstoftwo
 \else \expandafter \@secondoftwo
 \fi
}%
\providecommand \@ifx [1]{%
 \ifx #1\expandafter \@firstoftwo
 \else \expandafter \@secondoftwo
 \fi
}%
\providecommand \natexlab [1]{#1}%
\providecommand \enquote  [1]{``#1''}%
\providecommand \bibnamefont  [1]{#1}%
\providecommand \bibfnamefont [1]{#1}%
\providecommand \citenamefont [1]{#1}%
\providecommand \href@noop [0]{\@secondoftwo}%
\providecommand \href [0]{\begingroup \@sanitize@url \@href}%
\providecommand \@href[1]{\@@startlink{#1}\@@href}%
\providecommand \@@href[1]{\endgroup#1\@@endlink}%
\providecommand \@sanitize@url [0]{\catcode `\\12\catcode `\$12\catcode
  `\&12\catcode `\#12\catcode `\^12\catcode `\_12\catcode `\%12\relax}%
\providecommand \@@startlink[1]{}%
\providecommand \@@endlink[0]{}%
\providecommand \url  [0]{\begingroup\@sanitize@url \@url }%
\providecommand \@url [1]{\endgroup\@href {#1}{\urlprefix }}%
\providecommand \urlprefix  [0]{URL }%
\providecommand \Eprint [0]{\href }%
\providecommand \doibase [0]{http://dx.doi.org/}%
\providecommand \selectlanguage [0]{\@gobble}%
\providecommand \bibinfo  [0]{\@secondoftwo}%
\providecommand \bibfield  [0]{\@secondoftwo}%
\providecommand \translation [1]{[#1]}%
\providecommand \BibitemOpen [0]{}%
\providecommand \bibitemStop [0]{}%
\providecommand \bibitemNoStop [0]{.\EOS\space}%
\providecommand \EOS [0]{\spacefactor3000\relax}%
\providecommand \BibitemShut  [1]{\csname bibitem#1\endcsname}%
\let\auto@bib@innerbib\@empty
\bibitem [{\citenamefont {Militzer}\ \emph {et~al.}(2016)\citenamefont
  {Militzer}, \citenamefont {Soubiran}, \citenamefont {Wahl},\ and\
  \citenamefont {Hubbard}}]{militzer_understanding_2016}%
  \BibitemOpen
  \bibfield  {author} {\bibinfo {author} {\bibfnamefont {B.}~\bibnamefont
  {Militzer}}, \bibinfo {author} {\bibfnamefont {F.}~\bibnamefont {Soubiran}},
  \bibinfo {author} {\bibfnamefont {S.~M.}\ \bibnamefont {Wahl}}, \ and\
  \bibinfo {author} {\bibfnamefont {W.}~\bibnamefont {Hubbard}},\ }\href
  {\doibase 10.1002/2016JE005080} {\bibfield  {journal} {\bibinfo  {journal}
  {Journal of Geophysical Research: Planets}\ }\textbf {\bibinfo {volume}
  {121}},\ \bibinfo {pages} {1552} (\bibinfo {year} {2016})}\BibitemShut
  {NoStop}%
\bibitem [{\citenamefont {Mazzola}, \citenamefont {Helled},\ and\ \citenamefont
  {Sorella}(2018)}]{mazzola_phase_2018}%
  \BibitemOpen
  \bibfield  {author} {\bibinfo {author} {\bibfnamefont {G.}~\bibnamefont
  {Mazzola}}, \bibinfo {author} {\bibfnamefont {R.}~\bibnamefont {Helled}}, \
  and\ \bibinfo {author} {\bibfnamefont {S.}~\bibnamefont {Sorella}},\ }\href
  {\doibase 10.1103/PhysRevLett.120.025701} {\bibfield  {journal} {\bibinfo
  {journal} {Physical Review Letters}\ }\textbf {\bibinfo {volume} {120}},\
  \bibinfo {pages} {025701} (\bibinfo {year} {2018})}\BibitemShut {NoStop}%
\bibitem [{\citenamefont {Mukherjee}\ \emph {et~al.}(2013)\citenamefont
  {Mukherjee}, \citenamefont {Libisch}, \citenamefont {Large}, \citenamefont
  {Neumann}, \citenamefont {Brown}, \citenamefont {Cheng}, \citenamefont
  {Lassiter}, \citenamefont {Carter}, \citenamefont {Nordlander},\ and\
  \citenamefont {Halas}}]{mukherjee_hot_2013}%
  \BibitemOpen
  \bibfield  {author} {\bibinfo {author} {\bibfnamefont {S.}~\bibnamefont
  {Mukherjee}}, \bibinfo {author} {\bibfnamefont {F.}~\bibnamefont {Libisch}},
  \bibinfo {author} {\bibfnamefont {N.}~\bibnamefont {Large}}, \bibinfo
  {author} {\bibfnamefont {O.}~\bibnamefont {Neumann}}, \bibinfo {author}
  {\bibfnamefont {L.~V.}\ \bibnamefont {Brown}}, \bibinfo {author}
  {\bibfnamefont {J.}~\bibnamefont {Cheng}}, \bibinfo {author} {\bibfnamefont
  {J.~B.}\ \bibnamefont {Lassiter}}, \bibinfo {author} {\bibfnamefont {E.~A.}\
  \bibnamefont {Carter}}, \bibinfo {author} {\bibfnamefont {P.}~\bibnamefont
  {Nordlander}}, \ and\ \bibinfo {author} {\bibfnamefont {N.~J.}\ \bibnamefont
  {Halas}},\ }\href {\doibase 10.1021/nl303940z} {\bibfield  {journal}
  {\bibinfo  {journal} {Nano Letters}\ }\textbf {\bibinfo {volume} {13}},\
  \bibinfo {pages} {240} (\bibinfo {year} {2013})}\BibitemShut {NoStop}%
\bibitem [{\citenamefont {Zhou}\ \emph {et~al.}(2016)\citenamefont {Zhou},
  \citenamefont {Zhang}, \citenamefont {McClain}, \citenamefont {Manjavacas},
  \citenamefont {Krauter}, \citenamefont {Tian}, \citenamefont {Berg},
  \citenamefont {Everitt}, \citenamefont {Carter}, \citenamefont {Nordlander},\
  and\ \citenamefont {Halas}}]{zhou_aluminum_2016}%
  \BibitemOpen
  \bibfield  {author} {\bibinfo {author} {\bibfnamefont {L.}~\bibnamefont
  {Zhou}}, \bibinfo {author} {\bibfnamefont {C.}~\bibnamefont {Zhang}},
  \bibinfo {author} {\bibfnamefont {M.~J.}\ \bibnamefont {McClain}}, \bibinfo
  {author} {\bibfnamefont {A.}~\bibnamefont {Manjavacas}}, \bibinfo {author}
  {\bibfnamefont {C.~M.}\ \bibnamefont {Krauter}}, \bibinfo {author}
  {\bibfnamefont {S.}~\bibnamefont {Tian}}, \bibinfo {author} {\bibfnamefont
  {F.}~\bibnamefont {Berg}}, \bibinfo {author} {\bibfnamefont {H.~O.}\
  \bibnamefont {Everitt}}, \bibinfo {author} {\bibfnamefont {E.~A.}\
  \bibnamefont {Carter}}, \bibinfo {author} {\bibfnamefont {P.}~\bibnamefont
  {Nordlander}}, \ and\ \bibinfo {author} {\bibfnamefont {N.~J.}\ \bibnamefont
  {Halas}},\ }\href {\doibase 10.1021/acs.nanolett.5b05149} {\bibfield
  {journal} {\bibinfo  {journal} {Nano Letters}\ }\textbf {\bibinfo {volume}
  {16}},\ \bibinfo {pages} {1478} (\bibinfo {year} {2016})}\BibitemShut
  {NoStop}%
\bibitem [{\citenamefont {Ernstorfer}\ \emph {et~al.}(2009)\citenamefont
  {Ernstorfer}, \citenamefont {Harb}, \citenamefont {Hebeisen}, \citenamefont
  {Sciaini}, \citenamefont {Dartigalongue},\ and\ \citenamefont
  {Miller}}]{ernstorfer_formation_2009}%
  \BibitemOpen
  \bibfield  {author} {\bibinfo {author} {\bibfnamefont {R.}~\bibnamefont
  {Ernstorfer}}, \bibinfo {author} {\bibfnamefont {M.}~\bibnamefont {Harb}},
  \bibinfo {author} {\bibfnamefont {C.~T.}\ \bibnamefont {Hebeisen}}, \bibinfo
  {author} {\bibfnamefont {G.}~\bibnamefont {Sciaini}}, \bibinfo {author}
  {\bibfnamefont {T.}~\bibnamefont {Dartigalongue}}, \ and\ \bibinfo {author}
  {\bibfnamefont {R.~J.~D.}\ \bibnamefont {Miller}},\ }\href {\doibase
  10.1126/science.1162697} {\bibfield  {journal} {\bibinfo  {journal}
  {Science}\ }\textbf {\bibinfo {volume} {323}},\ \bibinfo {pages} {1033}
  (\bibinfo {year} {2009})}\BibitemShut {NoStop}%
\bibitem [{\citenamefont {Gull}, \citenamefont {Parcollet},\ and\ \citenamefont
  {Millis}(2013)}]{gull_superconductivity_2013}%
  \BibitemOpen
  \bibfield  {author} {\bibinfo {author} {\bibfnamefont {E.}~\bibnamefont
  {Gull}}, \bibinfo {author} {\bibfnamefont {O.}~\bibnamefont {Parcollet}}, \
  and\ \bibinfo {author} {\bibfnamefont {A.~J.}\ \bibnamefont {Millis}},\
  }\href {\doibase 10.1103/PhysRevLett.110.216405} {\bibfield  {journal}
  {\bibinfo  {journal} {Physical Review Letters}\ }\textbf {\bibinfo {volume}
  {110}},\ \bibinfo {pages} {216405} (\bibinfo {year} {2013})}\BibitemShut
  {NoStop}%
\bibitem [{\citenamefont {Drummond}\ \emph {et~al.}(2015)\citenamefont
  {Drummond}, \citenamefont {Monserrat}, \citenamefont {Lloyd-Williams},
  \citenamefont {Ríos}, \citenamefont {Pickard},\ and\ \citenamefont
  {Needs}}]{drummond_quantum_2015}%
  \BibitemOpen
  \bibfield  {author} {\bibinfo {author} {\bibfnamefont {N.~D.}\ \bibnamefont
  {Drummond}}, \bibinfo {author} {\bibfnamefont {B.}~\bibnamefont {Monserrat}},
  \bibinfo {author} {\bibfnamefont {J.~H.}\ \bibnamefont {Lloyd-Williams}},
  \bibinfo {author} {\bibfnamefont {P.~L.}\ \bibnamefont {Ríos}}, \bibinfo
  {author} {\bibfnamefont {C.~J.}\ \bibnamefont {Pickard}}, \ and\ \bibinfo
  {author} {\bibfnamefont {R.~J.}\ \bibnamefont {Needs}},\ }\href {\doibase
  10.1038/ncomms8794} {\bibfield  {journal} {\bibinfo  {journal} {Nature
  Communications}\ }\textbf {\bibinfo {volume} {6}},\ \bibinfo {pages} {7794}
  (\bibinfo {year} {2015})}\BibitemShut {NoStop}%
\bibitem [{\citenamefont {He}, \citenamefont {Ryu},\ and\ \citenamefont
  {Hirata}(2014)}]{he_finite-temperature_2014}%
  \BibitemOpen
  \bibfield  {author} {\bibinfo {author} {\bibfnamefont {X.}~\bibnamefont
  {He}}, \bibinfo {author} {\bibfnamefont {S.}~\bibnamefont {Ryu}}, \ and\
  \bibinfo {author} {\bibfnamefont {S.}~\bibnamefont {Hirata}},\ }\href
  {\doibase 10.1063/1.4859257} {\bibfield  {journal} {\bibinfo  {journal} {The
  Journal of Chemical Physics}\ }\textbf {\bibinfo {volume} {140}},\ \bibinfo
  {pages} {024702} (\bibinfo {year} {2014})}\BibitemShut {NoStop}%
\bibitem [{\citenamefont {Santra}\ and\ \citenamefont
  {Schirmer}(2017)}]{santra_finite-temperature_2017}%
  \BibitemOpen
  \bibfield  {author} {\bibinfo {author} {\bibfnamefont {R.}~\bibnamefont
  {Santra}}\ and\ \bibinfo {author} {\bibfnamefont {J.}~\bibnamefont
  {Schirmer}},\ }\href {\doibase 10.1016/j.chemphys.2016.08.001} {\bibfield
  {journal} {\bibinfo  {journal} {Chemical Physics}\ }\textbf {\bibinfo
  {volume} {482}},\ \bibinfo {pages} {355} (\bibinfo {year}
  {2017})}\BibitemShut {NoStop}%
\bibitem [{\citenamefont {Hirata}\ and\ \citenamefont
  {Jha}(2020)}]{hirata_finite-temperature_2020}%
  \BibitemOpen
  \bibfield  {author} {\bibinfo {author} {\bibfnamefont {S.}~\bibnamefont
  {Hirata}}\ and\ \bibinfo {author} {\bibfnamefont {P.~K.}\ \bibnamefont
  {Jha}},\ }\href {\doibase 10.1063/5.0009679} {\bibfield  {journal} {\bibinfo
  {journal} {The Journal of Chemical Physics}\ }\textbf {\bibinfo {volume}
  {153}},\ \bibinfo {pages} {014103} (\bibinfo {year} {2020})}\BibitemShut
  {NoStop}%
\bibitem [{\citenamefont {Jha}\ and\ \citenamefont
  {Hirata}(2020)}]{jha_finite-temperature_2020}%
  \BibitemOpen
  \bibfield  {author} {\bibinfo {author} {\bibfnamefont {P.~K.}\ \bibnamefont
  {Jha}}\ and\ \bibinfo {author} {\bibfnamefont {S.}~\bibnamefont {Hirata}},\
  }\href {\doibase 10.1103/PhysRevE.101.022106} {\bibfield  {journal} {\bibinfo
   {journal} {Physical Review E}\ }\textbf {\bibinfo {volume} {101}},\ \bibinfo
  {pages} {022106} (\bibinfo {year} {2020})}\BibitemShut {NoStop}%
\bibitem [{\citenamefont {Hirata}(2021)}]{hirata_finite-temperature_2021}%
  \BibitemOpen
  \bibfield  {author} {\bibinfo {author} {\bibfnamefont {S.}~\bibnamefont
  {Hirata}},\ }\href {\doibase 10.1063/5.0061384} {\bibfield  {journal}
  {\bibinfo  {journal} {The Journal of Chemical Physics}\ }\textbf {\bibinfo
  {volume} {155}},\ \bibinfo {pages} {094106} (\bibinfo {year}
  {2021})}\BibitemShut {NoStop}%
\bibitem [{\citenamefont {Dzhioev}\ and\ \citenamefont
  {Kosov}(2015)}]{dzhioev_superoperator_2015}%
  \BibitemOpen
  \bibfield  {author} {\bibinfo {author} {\bibfnamefont {A.~A.}\ \bibnamefont
  {Dzhioev}}\ and\ \bibinfo {author} {\bibfnamefont {D.~S.}\ \bibnamefont
  {Kosov}},\ }\href {\doibase 10.1088/1751-8113/48/1/015004} {\bibfield
  {journal} {\bibinfo  {journal} {Journal of Physics A: Mathematical and
  Theoretical}\ }\textbf {\bibinfo {volume} {48}},\ \bibinfo {pages} {015004}
  (\bibinfo {year} {2015})}\BibitemShut {NoStop}%
\bibitem [{\citenamefont {Hermes}\ and\ \citenamefont
  {Hirata}(2015)}]{hermes_finite-temperature_2015}%
  \BibitemOpen
  \bibfield  {author} {\bibinfo {author} {\bibfnamefont {M.~R.}\ \bibnamefont
  {Hermes}}\ and\ \bibinfo {author} {\bibfnamefont {S.}~\bibnamefont
  {Hirata}},\ }\href {\doibase 10.1063/1.4930024} {\bibfield  {journal}
  {\bibinfo  {journal} {The Journal of Chemical Physics}\ }\textbf {\bibinfo
  {volume} {143}},\ \bibinfo {pages} {102818} (\bibinfo {year}
  {2015})}\BibitemShut {NoStop}%
\bibitem [{\citenamefont {Hummel}(2018)}]{hummel_finite_2018}%
  \BibitemOpen
  \bibfield  {author} {\bibinfo {author} {\bibfnamefont {F.}~\bibnamefont
  {Hummel}},\ }\href {\doibase 10.1021/acs.jctc.8b00793} {\bibfield  {journal}
  {\bibinfo  {journal} {Journal of Chemical Theory and Computation}\ }\textbf
  {\bibinfo {volume} {14}},\ \bibinfo {pages} {6505} (\bibinfo {year}
  {2018})}\BibitemShut {NoStop}%
\bibitem [{\citenamefont {Harsha}, \citenamefont {Henderson},\ and\
  \citenamefont {Scuseria}(2019)}]{harsha_thermofield_2019}%
  \BibitemOpen
  \bibfield  {author} {\bibinfo {author} {\bibfnamefont {G.}~\bibnamefont
  {Harsha}}, \bibinfo {author} {\bibfnamefont {T.~M.}\ \bibnamefont
  {Henderson}}, \ and\ \bibinfo {author} {\bibfnamefont {G.~E.}\ \bibnamefont
  {Scuseria}},\ }\href {\doibase 10.1063/1.5089560} {\bibfield  {journal}
  {\bibinfo  {journal} {The Journal of Chemical Physics}\ }\textbf {\bibinfo
  {volume} {150}},\ \bibinfo {pages} {154109} (\bibinfo {year}
  {2019})}\BibitemShut {NoStop}%
\bibitem [{\citenamefont {White}\ and\ \citenamefont
  {Chan}(2019)}]{white_time-dependent_2019}%
  \BibitemOpen
  \bibfield  {author} {\bibinfo {author} {\bibfnamefont {A.~F.}\ \bibnamefont
  {White}}\ and\ \bibinfo {author} {\bibfnamefont {G.~K.-L.}\ \bibnamefont
  {Chan}},\ }\href {\doibase 10.1021/acs.jctc.9b00750} {\bibfield  {journal}
  {\bibinfo  {journal} {Journal of Chemical Theory and Computation}\ }\textbf
  {\bibinfo {volume} {15}},\ \bibinfo {pages} {6137} (\bibinfo {year}
  {2019})}\BibitemShut {NoStop}%
\bibitem [{\citenamefont {Shushkov}\ and\ \citenamefont
  {Miller}(2019)}]{shushkov_real-time_2019}%
  \BibitemOpen
  \bibfield  {author} {\bibinfo {author} {\bibfnamefont {P.}~\bibnamefont
  {Shushkov}}\ and\ \bibinfo {author} {\bibfnamefont {T.~F.}\ \bibnamefont
  {Miller}},\ }\href {\doibase 10.1063/1.5121749} {\bibfield  {journal}
  {\bibinfo  {journal} {The Journal of Chemical Physics}\ }\textbf {\bibinfo
  {volume} {151}},\ \bibinfo {pages} {134107} (\bibinfo {year}
  {2019})}\BibitemShut {NoStop}%
\bibitem [{\citenamefont {White}\ and\ \citenamefont
  {Kin-Lic~Chan}(2020)}]{white_finite-temperature_2020}%
  \BibitemOpen
  \bibfield  {author} {\bibinfo {author} {\bibfnamefont {A.~F.}\ \bibnamefont
  {White}}\ and\ \bibinfo {author} {\bibfnamefont {G.}~\bibnamefont
  {Kin-Lic~Chan}},\ }\href {\doibase 10.1063/5.0009845} {\bibfield  {journal}
  {\bibinfo  {journal} {The Journal of Chemical Physics}\ }\textbf {\bibinfo
  {volume} {152}},\ \bibinfo {pages} {224104} (\bibinfo {year}
  {2020})}\BibitemShut {NoStop}%
\bibitem [{\citenamefont {Peng}\ \emph {et~al.}(2021)\citenamefont {Peng},
  \citenamefont {White}, \citenamefont {Zhai},\ and\ \citenamefont
  {Kin-Lic~Chan}}]{peng_conservation_2021}%
  \BibitemOpen
  \bibfield  {author} {\bibinfo {author} {\bibfnamefont {R.}~\bibnamefont
  {Peng}}, \bibinfo {author} {\bibfnamefont {A.~F.}\ \bibnamefont {White}},
  \bibinfo {author} {\bibfnamefont {H.}~\bibnamefont {Zhai}}, \ and\ \bibinfo
  {author} {\bibfnamefont {G.}~\bibnamefont {Kin-Lic~Chan}},\ }\href {\doibase
  10.1063/5.0059257} {\bibfield  {journal} {\bibinfo  {journal} {The Journal of
  Chemical Physics}\ }\textbf {\bibinfo {volume} {155}},\ \bibinfo {pages}
  {044103} (\bibinfo {year} {2021})}\BibitemShut {NoStop}%
\bibitem [{\citenamefont {Harsha}\ \emph {et~al.}(2022)\citenamefont {Harsha},
  \citenamefont {Xu}, \citenamefont {Henderson},\ and\ \citenamefont
  {Scuseria}}]{harsha_thermal_2022}%
  \BibitemOpen
  \bibfield  {author} {\bibinfo {author} {\bibfnamefont {G.}~\bibnamefont
  {Harsha}}, \bibinfo {author} {\bibfnamefont {Y.}~\bibnamefont {Xu}}, \bibinfo
  {author} {\bibfnamefont {T.~M.}\ \bibnamefont {Henderson}}, \ and\ \bibinfo
  {author} {\bibfnamefont {G.~E.}\ \bibnamefont {Scuseria}},\ }\href {\doibase
  10.1103/PhysRevB.105.045125} {\bibfield  {journal} {\bibinfo  {journal}
  {Physical Review B}\ }\textbf {\bibinfo {volume} {105}},\ \bibinfo {pages}
  {045125} (\bibinfo {year} {2022})}\BibitemShut {NoStop}%
\bibitem [{\citenamefont {Karasiev}, \citenamefont {Sjostrom},\ and\
  \citenamefont
  {Trickey}(2012)}]{karasiev_generalized-gradient-approximation_2012}%
  \BibitemOpen
  \bibfield  {author} {\bibinfo {author} {\bibfnamefont {V.~V.}\ \bibnamefont
  {Karasiev}}, \bibinfo {author} {\bibfnamefont {T.}~\bibnamefont {Sjostrom}},
  \ and\ \bibinfo {author} {\bibfnamefont {S.~B.}\ \bibnamefont {Trickey}},\
  }\href {\doibase 10.1103/PhysRevB.86.115101} {\bibfield  {journal} {\bibinfo
  {journal} {Physical Review B}\ }\textbf {\bibinfo {volume} {86}},\ \bibinfo
  {pages} {115101} (\bibinfo {year} {2012})}\BibitemShut {NoStop}%
\bibitem [{\citenamefont {Ellis}\ \emph {et~al.}(2021)\citenamefont {Ellis},
  \citenamefont {Fiedler}, \citenamefont {Popoola}, \citenamefont {Modine},
  \citenamefont {Stephens}, \citenamefont {Thompson}, \citenamefont {Cangi},\
  and\ \citenamefont {Rajamanickam}}]{ellis_accelerating_2021}%
  \BibitemOpen
  \bibfield  {author} {\bibinfo {author} {\bibfnamefont {J.~A.}\ \bibnamefont
  {Ellis}}, \bibinfo {author} {\bibfnamefont {L.}~\bibnamefont {Fiedler}},
  \bibinfo {author} {\bibfnamefont {G.~A.}\ \bibnamefont {Popoola}}, \bibinfo
  {author} {\bibfnamefont {N.~A.}\ \bibnamefont {Modine}}, \bibinfo {author}
  {\bibfnamefont {J.~A.}\ \bibnamefont {Stephens}}, \bibinfo {author}
  {\bibfnamefont {A.~P.}\ \bibnamefont {Thompson}}, \bibinfo {author}
  {\bibfnamefont {A.}~\bibnamefont {Cangi}}, \ and\ \bibinfo {author}
  {\bibfnamefont {S.}~\bibnamefont {Rajamanickam}},\ }\href {\doibase
  10.1103/PhysRevB.104.035120} {\bibfield  {journal} {\bibinfo  {journal}
  {Physical Review B}\ }\textbf {\bibinfo {volume} {104}},\ \bibinfo {pages}
  {035120} (\bibinfo {year} {2021})}\BibitemShut {NoStop}%
\bibitem [{\citenamefont {Pittalis}\ \emph {et~al.}(2011)\citenamefont
  {Pittalis}, \citenamefont {Proetto}, \citenamefont {Floris}, \citenamefont
  {Sanna}, \citenamefont {Bersier}, \citenamefont {Burke},\ and\ \citenamefont
  {Gross}}]{pittalis_exact_2011}%
  \BibitemOpen
  \bibfield  {author} {\bibinfo {author} {\bibfnamefont {S.}~\bibnamefont
  {Pittalis}}, \bibinfo {author} {\bibfnamefont {C.~R.}\ \bibnamefont
  {Proetto}}, \bibinfo {author} {\bibfnamefont {A.}~\bibnamefont {Floris}},
  \bibinfo {author} {\bibfnamefont {A.}~\bibnamefont {Sanna}}, \bibinfo
  {author} {\bibfnamefont {C.}~\bibnamefont {Bersier}}, \bibinfo {author}
  {\bibfnamefont {K.}~\bibnamefont {Burke}}, \ and\ \bibinfo {author}
  {\bibfnamefont {E.~K.~U.}\ \bibnamefont {Gross}},\ }\href {\doibase
  10.1103/PhysRevLett.107.163001} {\bibfield  {journal} {\bibinfo  {journal}
  {Physical Review Letters}\ }\textbf {\bibinfo {volume} {107}},\ \bibinfo
  {pages} {163001} (\bibinfo {year} {2011})}\BibitemShut {NoStop}%
\bibitem [{\citenamefont {Eschrig}(2010)}]{eschrig_t_2010}%
  \BibitemOpen
  \bibfield  {author} {\bibinfo {author} {\bibfnamefont {H.}~\bibnamefont
  {Eschrig}},\ }\href {\doibase 10.1103/PhysRevB.82.205120} {\bibfield
  {journal} {\bibinfo  {journal} {Physical Review B}\ }\textbf {\bibinfo
  {volume} {82}},\ \bibinfo {pages} {205120} (\bibinfo {year}
  {2010})}\BibitemShut {NoStop}%
\bibitem [{\citenamefont {Pribram-Jones}, \citenamefont {Grabowski},\ and\
  \citenamefont {Burke}(2016)}]{pribram-jones_thermal_2016}%
  \BibitemOpen
  \bibfield  {author} {\bibinfo {author} {\bibfnamefont {A.}~\bibnamefont
  {Pribram-Jones}}, \bibinfo {author} {\bibfnamefont {P.~E.}\ \bibnamefont
  {Grabowski}}, \ and\ \bibinfo {author} {\bibfnamefont {K.}~\bibnamefont
  {Burke}},\ }\href {\doibase 10.1103/PhysRevLett.116.233001} {\bibfield
  {journal} {\bibinfo  {journal} {Physical Review Letters}\ }\textbf {\bibinfo
  {volume} {116}},\ \bibinfo {pages} {233001} (\bibinfo {year}
  {2016})}\BibitemShut {NoStop}%
\bibitem [{\citenamefont {Kananenka}\ \emph {et~al.}(2016)\citenamefont
  {Kananenka}, \citenamefont {Welden}, \citenamefont {Lan}, \citenamefont
  {Gull},\ and\ \citenamefont {Zgid}}]{kananenka_efficient_2016}%
  \BibitemOpen
  \bibfield  {author} {\bibinfo {author} {\bibfnamefont {A.~A.}\ \bibnamefont
  {Kananenka}}, \bibinfo {author} {\bibfnamefont {A.~R.}\ \bibnamefont
  {Welden}}, \bibinfo {author} {\bibfnamefont {T.~N.}\ \bibnamefont {Lan}},
  \bibinfo {author} {\bibfnamefont {E.}~\bibnamefont {Gull}}, \ and\ \bibinfo
  {author} {\bibfnamefont {D.}~\bibnamefont {Zgid}},\ }\href {\doibase
  10.1021/acs.jctc.6b00178} {\bibfield  {journal} {\bibinfo  {journal} {Journal
  of Chemical Theory and Computation}\ }\textbf {\bibinfo {volume} {12}},\
  \bibinfo {pages} {2250} (\bibinfo {year} {2016})}\BibitemShut {NoStop}%
\bibitem [{\citenamefont {Welden}, \citenamefont {Rusakov},\ and\ \citenamefont
  {Zgid}(2016)}]{welden_exploring_2016}%
  \BibitemOpen
  \bibfield  {author} {\bibinfo {author} {\bibfnamefont {A.~R.}\ \bibnamefont
  {Welden}}, \bibinfo {author} {\bibfnamefont {A.~A.}\ \bibnamefont {Rusakov}},
  \ and\ \bibinfo {author} {\bibfnamefont {D.}~\bibnamefont {Zgid}},\ }\href
  {\doibase 10.1063/1.4967449} {\bibfield  {journal} {\bibinfo  {journal} {The
  Journal of Chemical Physics}\ }\textbf {\bibinfo {volume} {145}},\ \bibinfo
  {pages} {204106} (\bibinfo {year} {2016})}\BibitemShut {NoStop}%
\bibitem [{\citenamefont {Kas}\ and\ \citenamefont
  {Rehr}(2017)}]{kas_finite_2017}%
  \BibitemOpen
  \bibfield  {author} {\bibinfo {author} {\bibfnamefont {J.}~\bibnamefont
  {Kas}}\ and\ \bibinfo {author} {\bibfnamefont {J.}~\bibnamefont {Rehr}},\
  }\href {\doibase 10.1103/PhysRevLett.119.176403} {\bibfield  {journal}
  {\bibinfo  {journal} {Physical Review Letters}\ }\textbf {\bibinfo {volume}
  {119}},\ \bibinfo {pages} {176403} (\bibinfo {year} {2017})}\BibitemShut
  {NoStop}%
\bibitem [{\citenamefont {Karrasch}, \citenamefont {Meden},\ and\ \citenamefont
  {Schönhammer}(2010)}]{karrasch_finite-temperature_2010}%
  \BibitemOpen
  \bibfield  {author} {\bibinfo {author} {\bibfnamefont {C.}~\bibnamefont
  {Karrasch}}, \bibinfo {author} {\bibfnamefont {V.}~\bibnamefont {Meden}}, \
  and\ \bibinfo {author} {\bibfnamefont {K.}~\bibnamefont {Schönhammer}},\
  }\href {\doibase 10.1103/PhysRevB.82.125114} {\bibfield  {journal} {\bibinfo
  {journal} {Physical Review B}\ }\textbf {\bibinfo {volume} {82}},\ \bibinfo
  {pages} {125114} (\bibinfo {year} {2010})}\BibitemShut {NoStop}%
\bibitem [{\citenamefont {Neuhauser}, \citenamefont {Baer},\ and\ \citenamefont
  {Zgid}(2017)}]{neuhauser_stochastic_2017}%
  \BibitemOpen
  \bibfield  {author} {\bibinfo {author} {\bibfnamefont {D.}~\bibnamefont
  {Neuhauser}}, \bibinfo {author} {\bibfnamefont {R.}~\bibnamefont {Baer}}, \
  and\ \bibinfo {author} {\bibfnamefont {D.}~\bibnamefont {Zgid}},\ }\href
  {\doibase 10.1021/acs.jctc.7b00792} {\bibfield  {journal} {\bibinfo
  {journal} {Journal of Chemical Theory and Computation}\ }\textbf {\bibinfo
  {volume} {13}},\ \bibinfo {pages} {5396} (\bibinfo {year}
  {2017})}\BibitemShut {NoStop}%
\bibitem [{\citenamefont {Gu}\ \emph {et~al.}(2020)\citenamefont {Gu},
  \citenamefont {Chen}, \citenamefont {Wang},\ and\ \citenamefont
  {Zhang}}]{gu_generalized_2020}%
  \BibitemOpen
  \bibfield  {author} {\bibinfo {author} {\bibfnamefont {J.}~\bibnamefont
  {Gu}}, \bibinfo {author} {\bibfnamefont {J.}~\bibnamefont {Chen}}, \bibinfo
  {author} {\bibfnamefont {Y.}~\bibnamefont {Wang}}, \ and\ \bibinfo {author}
  {\bibfnamefont {X.-G.}\ \bibnamefont {Zhang}},\ }\href {\doibase
  10.1016/j.cpc.2020.107178} {\bibfield  {journal} {\bibinfo  {journal}
  {Computer Physics Communications}\ }\textbf {\bibinfo {volume} {253}},\
  \bibinfo {pages} {107178} (\bibinfo {year} {2020})}\BibitemShut {NoStop}%
\bibitem [{\citenamefont {Li}\ \emph {et~al.}(2020)\citenamefont {Li},
  \citenamefont {Wallerberger}, \citenamefont {Chikano}, \citenamefont {Yeh},
  \citenamefont {Gull},\ and\ \citenamefont {Shinaoka}}]{li_sparse_2020}%
  \BibitemOpen
  \bibfield  {author} {\bibinfo {author} {\bibfnamefont {J.}~\bibnamefont
  {Li}}, \bibinfo {author} {\bibfnamefont {M.}~\bibnamefont {Wallerberger}},
  \bibinfo {author} {\bibfnamefont {N.}~\bibnamefont {Chikano}}, \bibinfo
  {author} {\bibfnamefont {C.-N.}\ \bibnamefont {Yeh}}, \bibinfo {author}
  {\bibfnamefont {E.}~\bibnamefont {Gull}}, \ and\ \bibinfo {author}
  {\bibfnamefont {H.}~\bibnamefont {Shinaoka}},\ }\href {\doibase
  10.1103/PhysRevB.101.035144} {\bibfield  {journal} {\bibinfo  {journal}
  {Physical Review B}\ }\textbf {\bibinfo {volume} {101}},\ \bibinfo {pages}
  {035144} (\bibinfo {year} {2020})}\BibitemShut {NoStop}%
\bibitem [{\citenamefont {Knizia}\ and\ \citenamefont
  {Chan}(2013)}]{knizia_density_2013}%
  \BibitemOpen
  \bibfield  {author} {\bibinfo {author} {\bibfnamefont {G.}~\bibnamefont
  {Knizia}}\ and\ \bibinfo {author} {\bibfnamefont {G.~K.-L.}\ \bibnamefont
  {Chan}},\ }\href {\doibase 10.1021/ct301044e} {\bibfield  {journal} {\bibinfo
   {journal} {Journal of Chemical Theory and Computation}\ }\textbf {\bibinfo
  {volume} {9}},\ \bibinfo {pages} {1428} (\bibinfo {year} {2013})}\BibitemShut
  {NoStop}%
\bibitem [{\citenamefont {Sun}\ \emph {et~al.}(2020)\citenamefont {Sun},
  \citenamefont {Ray}, \citenamefont {Cui}, \citenamefont {Stoudenmire},
  \citenamefont {Ferrero},\ and\ \citenamefont
  {Chan}}]{sun_finite-temperature_2020}%
  \BibitemOpen
  \bibfield  {author} {\bibinfo {author} {\bibfnamefont {C.}~\bibnamefont
  {Sun}}, \bibinfo {author} {\bibfnamefont {U.}~\bibnamefont {Ray}}, \bibinfo
  {author} {\bibfnamefont {Z.-H.}\ \bibnamefont {Cui}}, \bibinfo {author}
  {\bibfnamefont {M.}~\bibnamefont {Stoudenmire}}, \bibinfo {author}
  {\bibfnamefont {M.}~\bibnamefont {Ferrero}}, \ and\ \bibinfo {author}
  {\bibfnamefont {G.~K.-L.}\ \bibnamefont {Chan}},\ }\href {\doibase
  10.1103/PhysRevB.101.075131} {\bibfield  {journal} {\bibinfo  {journal}
  {Physical Review B}\ }\textbf {\bibinfo {volume} {101}},\ \bibinfo {pages}
  {075131} (\bibinfo {year} {2020})}\BibitemShut {NoStop}%
\bibitem [{\citenamefont {Kretchmer}\ and\ \citenamefont
  {Chan}(2018)}]{kretchmer_real-time_2018}%
  \BibitemOpen
  \bibfield  {author} {\bibinfo {author} {\bibfnamefont {J.~S.}\ \bibnamefont
  {Kretchmer}}\ and\ \bibinfo {author} {\bibfnamefont {G.~K.-L.}\ \bibnamefont
  {Chan}},\ }\href {\doibase 10.1063/1.5012766} {\bibfield  {journal} {\bibinfo
   {journal} {The Journal of Chemical Physics}\ }\textbf {\bibinfo {volume}
  {148}},\ \bibinfo {pages} {054108} (\bibinfo {year} {2018})}\BibitemShut
  {NoStop}%
\bibitem [{\citenamefont {Tran}, \citenamefont {Van~Voorhis},\ and\
  \citenamefont {Thom}(2019)}]{tran_using_2019}%
  \BibitemOpen
  \bibfield  {author} {\bibinfo {author} {\bibfnamefont {H.~K.}\ \bibnamefont
  {Tran}}, \bibinfo {author} {\bibfnamefont {T.}~\bibnamefont {Van~Voorhis}}, \
  and\ \bibinfo {author} {\bibfnamefont {A.~J.~W.}\ \bibnamefont {Thom}},\
  }\href {\doibase 10.1063/1.5096177} {\bibfield  {journal} {\bibinfo
  {journal} {The Journal of Chemical Physics}\ }\textbf {\bibinfo {volume}
  {151}},\ \bibinfo {pages} {034112} (\bibinfo {year} {2019})}\BibitemShut
  {NoStop}%
\bibitem [{\citenamefont {Cui}, \citenamefont {Zhu},\ and\ \citenamefont
  {Chan}(2020)}]{cui_efficient_2020}%
  \BibitemOpen
  \bibfield  {author} {\bibinfo {author} {\bibfnamefont {Z.-H.}\ \bibnamefont
  {Cui}}, \bibinfo {author} {\bibfnamefont {T.}~\bibnamefont {Zhu}}, \ and\
  \bibinfo {author} {\bibfnamefont {G.~K.-L.}\ \bibnamefont {Chan}},\ }\href
  {\doibase 10.1021/acs.jctc.9b00933} {\bibfield  {journal} {\bibinfo
  {journal} {Journal of Chemical Theory and Computation}\ }\textbf {\bibinfo
  {volume} {16}},\ \bibinfo {pages} {119} (\bibinfo {year} {2020})}\BibitemShut
  {NoStop}%
\bibitem [{\citenamefont {Zhai}\ and\ \citenamefont
  {Chan}(2021)}]{zhai_low_2021}%
  \BibitemOpen
  \bibfield  {author} {\bibinfo {author} {\bibfnamefont {H.}~\bibnamefont
  {Zhai}}\ and\ \bibinfo {author} {\bibfnamefont {G.~K.-L.}\ \bibnamefont
  {Chan}},\ }\href {\doibase 10.1063/5.0050902} {\bibfield  {journal} {\bibinfo
   {journal} {The Journal of Chemical Physics}\ }\textbf {\bibinfo {volume}
  {154}},\ \bibinfo {pages} {224116} (\bibinfo {year} {2021})}\BibitemShut
  {NoStop}%
\bibitem [{\citenamefont {Tsuchimochi}, \citenamefont {Welborn},\ and\
  \citenamefont {Van~Voorhis}(2015)}]{tsuchimochi_density_2015}%
  \BibitemOpen
  \bibfield  {author} {\bibinfo {author} {\bibfnamefont {T.}~\bibnamefont
  {Tsuchimochi}}, \bibinfo {author} {\bibfnamefont {M.}~\bibnamefont
  {Welborn}}, \ and\ \bibinfo {author} {\bibfnamefont {T.}~\bibnamefont
  {Van~Voorhis}},\ }\href {\doibase 10.1063/1.4926650} {\bibfield  {journal}
  {\bibinfo  {journal} {The Journal of Chemical Physics}\ }\textbf {\bibinfo
  {volume} {143}},\ \bibinfo {pages} {024107} (\bibinfo {year}
  {2015})}\BibitemShut {NoStop}%
\bibitem [{\citenamefont {Bulik}, \citenamefont {Chen},\ and\ \citenamefont
  {Scuseria}(2014)}]{bulik_electron_2014}%
  \BibitemOpen
  \bibfield  {author} {\bibinfo {author} {\bibfnamefont {I.~W.}\ \bibnamefont
  {Bulik}}, \bibinfo {author} {\bibfnamefont {W.}~\bibnamefont {Chen}}, \ and\
  \bibinfo {author} {\bibfnamefont {G.~E.}\ \bibnamefont {Scuseria}},\ }\href
  {\doibase 10.1063/1.4891861} {\bibfield  {journal} {\bibinfo  {journal} {The
  Journal of Chemical Physics}\ }\textbf {\bibinfo {volume} {141}},\ \bibinfo
  {pages} {054113} (\bibinfo {year} {2014})}\BibitemShut {NoStop}%
\bibitem [{\citenamefont {Hermes}\ and\ \citenamefont
  {Gagliardi}(2019)}]{hermes_multiconfigurational_2019}%
  \BibitemOpen
  \bibfield  {author} {\bibinfo {author} {\bibfnamefont {M.~R.}\ \bibnamefont
  {Hermes}}\ and\ \bibinfo {author} {\bibfnamefont {L.}~\bibnamefont
  {Gagliardi}},\ }\href {\doibase 10.1021/acs.jctc.8b01009} {\bibfield
  {journal} {\bibinfo  {journal} {Journal of Chemical Theory and Computation}\
  }\textbf {\bibinfo {volume} {15}},\ \bibinfo {pages} {972} (\bibinfo {year}
  {2019})}\BibitemShut {NoStop}%
\bibitem [{\citenamefont {Zgid}\ and\ \citenamefont
  {Gull}(2017)}]{zgid_finite_2017}%
  \BibitemOpen
  \bibfield  {author} {\bibinfo {author} {\bibfnamefont {D.}~\bibnamefont
  {Zgid}}\ and\ \bibinfo {author} {\bibfnamefont {E.}~\bibnamefont {Gull}},\
  }\href {\doibase 10.1088/1367-2630/aa5d34} {\bibfield  {journal} {\bibinfo
  {journal} {New Journal of Physics}\ }\textbf {\bibinfo {volume} {19}},\
  \bibinfo {pages} {023047} (\bibinfo {year} {2017})}\BibitemShut {NoStop}%
\bibitem [{\citenamefont {Lan}\ and\ \citenamefont
  {Zgid}(2017)}]{lan_generalized_2017}%
  \BibitemOpen
  \bibfield  {author} {\bibinfo {author} {\bibfnamefont {T.~N.}\ \bibnamefont
  {Lan}}\ and\ \bibinfo {author} {\bibfnamefont {D.}~\bibnamefont {Zgid}},\
  }\href {\doibase 10.1021/acs.jpclett.7b00689} {\bibfield  {journal} {\bibinfo
   {journal} {The Journal of Physical Chemistry Letters}\ }\textbf {\bibinfo
  {volume} {8}},\ \bibinfo {pages} {2200} (\bibinfo {year} {2017})}\BibitemShut
  {NoStop}%
\bibitem [{\citenamefont {Tran}, \citenamefont {Iskakov},\ and\ \citenamefont
  {Zgid}(2018)}]{tran_spin-unrestricted_2018}%
  \BibitemOpen
  \bibfield  {author} {\bibinfo {author} {\bibfnamefont {L.~N.}\ \bibnamefont
  {Tran}}, \bibinfo {author} {\bibfnamefont {S.}~\bibnamefont {Iskakov}}, \
  and\ \bibinfo {author} {\bibfnamefont {D.}~\bibnamefont {Zgid}},\ }\href
  {\doibase 10.1021/acs.jpclett.8b01754} {\bibfield  {journal} {\bibinfo
  {journal} {The Journal of Physical Chemistry Letters}\ }\textbf {\bibinfo
  {volume} {9}},\ \bibinfo {pages} {4444} (\bibinfo {year} {2018})}\BibitemShut
  {NoStop}%
\bibitem [{\citenamefont {Rusakov}\ \emph {et~al.}(2019)\citenamefont
  {Rusakov}, \citenamefont {Iskakov}, \citenamefont {Tran},\ and\ \citenamefont
  {Zgid}}]{rusakov_self-energy_2019}%
  \BibitemOpen
  \bibfield  {author} {\bibinfo {author} {\bibfnamefont {A.~A.}\ \bibnamefont
  {Rusakov}}, \bibinfo {author} {\bibfnamefont {S.}~\bibnamefont {Iskakov}},
  \bibinfo {author} {\bibfnamefont {L.~N.}\ \bibnamefont {Tran}}, \ and\
  \bibinfo {author} {\bibfnamefont {D.}~\bibnamefont {Zgid}},\ }\href {\doibase
  10.1021/acs.jctc.8b00927} {\bibfield  {journal} {\bibinfo  {journal} {Journal
  of Chemical Theory and Computation}\ }\textbf {\bibinfo {volume} {15}},\
  \bibinfo {pages} {229} (\bibinfo {year} {2019})}\BibitemShut {NoStop}%
\bibitem [{\citenamefont {Dornheim}\ \emph {et~al.}(2015)\citenamefont
  {Dornheim}, \citenamefont {Groth}, \citenamefont {Filinov},\ and\
  \citenamefont {Bonitz}}]{dornheim_permutation_2015}%
  \BibitemOpen
  \bibfield  {author} {\bibinfo {author} {\bibfnamefont {T.}~\bibnamefont
  {Dornheim}}, \bibinfo {author} {\bibfnamefont {S.}~\bibnamefont {Groth}},
  \bibinfo {author} {\bibfnamefont {A.}~\bibnamefont {Filinov}}, \ and\
  \bibinfo {author} {\bibfnamefont {M.}~\bibnamefont {Bonitz}},\ }\href
  {\doibase 10.1088/1367-2630/17/7/073017} {\bibfield  {journal} {\bibinfo
  {journal} {New Journal of Physics}\ }\textbf {\bibinfo {volume} {17}},\
  \bibinfo {pages} {073017} (\bibinfo {year} {2015})}\BibitemShut {NoStop}%
\bibitem [{\citenamefont {Militzer}\ and\ \citenamefont
  {Driver}(2015)}]{militzer_development_2015}%
  \BibitemOpen
  \bibfield  {author} {\bibinfo {author} {\bibfnamefont {B.}~\bibnamefont
  {Militzer}}\ and\ \bibinfo {author} {\bibfnamefont {K.~P.}\ \bibnamefont
  {Driver}},\ }\href {\doibase 10.1103/PhysRevLett.115.176403} {\bibfield
  {journal} {\bibinfo  {journal} {Physical Review Letters}\ }\textbf {\bibinfo
  {volume} {115}},\ \bibinfo {pages} {176403} (\bibinfo {year}
  {2015})}\BibitemShut {NoStop}%
\bibitem [{\citenamefont {Larkin}\ and\ \citenamefont
  {Filinov}(2017)}]{larkin_phase_2017}%
  \BibitemOpen
  \bibfield  {author} {\bibinfo {author} {\bibfnamefont {A.~S.}\ \bibnamefont
  {Larkin}}\ and\ \bibinfo {author} {\bibfnamefont {V.~S.}\ \bibnamefont
  {Filinov}},\ }\href {\doibase 10.4236/jamp.2017.52035} {\bibfield  {journal}
  {\bibinfo  {journal} {Journal of Applied Mathematics and Physics}\ }\textbf
  {\bibinfo {volume} {05}},\ \bibinfo {pages} {392} (\bibinfo {year}
  {2017})}\BibitemShut {NoStop}%
\bibitem [{\citenamefont {Groth}, \citenamefont {Dornheim},\ and\ \citenamefont
  {Bonitz}(2017)}]{groth_configuration_2017}%
  \BibitemOpen
  \bibfield  {author} {\bibinfo {author} {\bibfnamefont {S.}~\bibnamefont
  {Groth}}, \bibinfo {author} {\bibfnamefont {T.}~\bibnamefont {Dornheim}}, \
  and\ \bibinfo {author} {\bibfnamefont {M.}~\bibnamefont {Bonitz}},\ }\href
  {\doibase 10.1063/1.4999907} {\bibfield  {journal} {\bibinfo  {journal} {The
  Journal of Chemical Physics}\ }\textbf {\bibinfo {volume} {147}},\ \bibinfo
  {pages} {164108} (\bibinfo {year} {2017})}\BibitemShut {NoStop}%
\bibitem [{\citenamefont {Dornheim}\ \emph {et~al.}(2018)\citenamefont
  {Dornheim}, \citenamefont {Groth}, \citenamefont {Vorberger},\ and\
  \citenamefont {Bonitz}}]{dornheim_ab_2018}%
  \BibitemOpen
  \bibfield  {author} {\bibinfo {author} {\bibfnamefont {T.}~\bibnamefont
  {Dornheim}}, \bibinfo {author} {\bibfnamefont {S.}~\bibnamefont {Groth}},
  \bibinfo {author} {\bibfnamefont {J.}~\bibnamefont {Vorberger}}, \ and\
  \bibinfo {author} {\bibfnamefont {M.}~\bibnamefont {Bonitz}},\ }\href
  {\doibase 10.1103/PhysRevLett.121.255001} {\bibfield  {journal} {\bibinfo
  {journal} {Physical Review Letters}\ }\textbf {\bibinfo {volume} {121}},\
  \bibinfo {pages} {255001} (\bibinfo {year} {2018})}\BibitemShut {NoStop}%
\bibitem [{\citenamefont {Dornheim}(2019)}]{dornheim_fermion_2019}%
  \BibitemOpen
  \bibfield  {author} {\bibinfo {author} {\bibfnamefont {T.}~\bibnamefont
  {Dornheim}},\ }\href {\doibase 10.1103/PhysRevE.100.023307} {\bibfield
  {journal} {\bibinfo  {journal} {Physical Review E}\ }\textbf {\bibinfo
  {volume} {100}},\ \bibinfo {pages} {023307} (\bibinfo {year}
  {2019})}\BibitemShut {NoStop}%
\bibitem [{\citenamefont {Yilmaz}\ \emph {et~al.}(2020)\citenamefont {Yilmaz},
  \citenamefont {Hunger}, \citenamefont {Dornheim}, \citenamefont {Groth},\
  and\ \citenamefont {Bonitz}}]{yilmaz_restricted_2020}%
  \BibitemOpen
  \bibfield  {author} {\bibinfo {author} {\bibfnamefont {A.}~\bibnamefont
  {Yilmaz}}, \bibinfo {author} {\bibfnamefont {K.}~\bibnamefont {Hunger}},
  \bibinfo {author} {\bibfnamefont {T.}~\bibnamefont {Dornheim}}, \bibinfo
  {author} {\bibfnamefont {S.}~\bibnamefont {Groth}}, \ and\ \bibinfo {author}
  {\bibfnamefont {M.}~\bibnamefont {Bonitz}},\ }\href {\doibase
  10.1063/5.0022800} {\bibfield  {journal} {\bibinfo  {journal} {The Journal of
  Chemical Physics}\ }\textbf {\bibinfo {volume} {153}},\ \bibinfo {pages}
  {124114} (\bibinfo {year} {2020})}\BibitemShut {NoStop}%
\bibitem [{\citenamefont {Dornheim}\ \emph {et~al.}(2021)\citenamefont
  {Dornheim}, \citenamefont {Böhme}, \citenamefont {Militzer},\ and\
  \citenamefont {Vorberger}}]{dornheim_ab_2021}%
  \BibitemOpen
  \bibfield  {author} {\bibinfo {author} {\bibfnamefont {T.}~\bibnamefont
  {Dornheim}}, \bibinfo {author} {\bibfnamefont {M.}~\bibnamefont {Böhme}},
  \bibinfo {author} {\bibfnamefont {B.}~\bibnamefont {Militzer}}, \ and\
  \bibinfo {author} {\bibfnamefont {J.}~\bibnamefont {Vorberger}},\ }\href
  {\doibase 10.1103/PhysRevB.103.205142} {\bibfield  {journal} {\bibinfo
  {journal} {Physical Review B}\ }\textbf {\bibinfo {volume} {103}},\ \bibinfo
  {pages} {205142} (\bibinfo {year} {2021})}\BibitemShut {NoStop}%
\bibitem [{\citenamefont {Lee}, \citenamefont {Chen},\ and\ \citenamefont
  {Kao}(2012)}]{lee_parallelizing_2012}%
  \BibitemOpen
  \bibfield  {author} {\bibinfo {author} {\bibfnamefont {C.-R.}\ \bibnamefont
  {Lee}}, \bibinfo {author} {\bibfnamefont {Z.-H.}\ \bibnamefont {Chen}}, \
  and\ \bibinfo {author} {\bibfnamefont {Q.-L.}\ \bibnamefont {Kao}},\ }in\
  \href {\doibase 10.1109/IPDPSW.2012.233} {\emph {\bibinfo {booktitle} {2012
  {IEEE} 26th {International} {Parallel} and {Distributed} {Processing}
  {Symposium} {Workshops} \& {PhD} {Forum}}}}\ (\bibinfo  {publisher} {IEEE},\
  \bibinfo {address} {Shanghai, China},\ \bibinfo {year} {2012})\ pp.\ \bibinfo
  {pages} {1889--1897}\BibitemShut {NoStop}%
\bibitem [{\citenamefont {Chang}\ \emph {et~al.}(2015)\citenamefont {Chang},
  \citenamefont {Gogolenko}, \citenamefont {Perez}, \citenamefont {Bai},\ and\
  \citenamefont {Scalettar}}]{chang_recent_2015}%
  \BibitemOpen
  \bibfield  {author} {\bibinfo {author} {\bibfnamefont {C.-C.}\ \bibnamefont
  {Chang}}, \bibinfo {author} {\bibfnamefont {S.}~\bibnamefont {Gogolenko}},
  \bibinfo {author} {\bibfnamefont {J.}~\bibnamefont {Perez}}, \bibinfo
  {author} {\bibfnamefont {Z.}~\bibnamefont {Bai}}, \ and\ \bibinfo {author}
  {\bibfnamefont {R.~T.}\ \bibnamefont {Scalettar}},\ }\href {\doibase
  10.1080/14786435.2013.845314} {\bibfield  {journal} {\bibinfo  {journal}
  {Philosophical Magazine}\ }\textbf {\bibinfo {volume} {95}},\ \bibinfo
  {pages} {1260} (\bibinfo {year} {2015})}\BibitemShut {NoStop}%
\bibitem [{\citenamefont {Liu}, \citenamefont {Cho},\ and\ \citenamefont
  {Rubenstein}(2018)}]{liu_ab_2018}%
  \BibitemOpen
  \bibfield  {author} {\bibinfo {author} {\bibfnamefont {Y.}~\bibnamefont
  {Liu}}, \bibinfo {author} {\bibfnamefont {M.}~\bibnamefont {Cho}}, \ and\
  \bibinfo {author} {\bibfnamefont {B.}~\bibnamefont {Rubenstein}},\ }\href
  {\doibase 10.1021/acs.jctc.8b00569} {\bibfield  {journal} {\bibinfo
  {journal} {Journal of Chemical Theory and Computation}\ }\textbf {\bibinfo
  {volume} {14}},\ \bibinfo {pages} {4722} (\bibinfo {year}
  {2018})}\BibitemShut {NoStop}%
\bibitem [{\citenamefont {He}\ \emph {et~al.}(2019)\citenamefont {He},
  \citenamefont {Qin}, \citenamefont {Shi}, \citenamefont {Lu},\ and\
  \citenamefont {Zhang}}]{he_finite-temperature_2019}%
  \BibitemOpen
  \bibfield  {author} {\bibinfo {author} {\bibfnamefont {Y.-Y.}\ \bibnamefont
  {He}}, \bibinfo {author} {\bibfnamefont {M.}~\bibnamefont {Qin}}, \bibinfo
  {author} {\bibfnamefont {H.}~\bibnamefont {Shi}}, \bibinfo {author}
  {\bibfnamefont {Z.-Y.}\ \bibnamefont {Lu}}, \ and\ \bibinfo {author}
  {\bibfnamefont {S.}~\bibnamefont {Zhang}},\ }\href {\doibase
  10.1103/PhysRevB.99.045108} {\bibfield  {journal} {\bibinfo  {journal}
  {Physical Review B}\ }\textbf {\bibinfo {volume} {99}},\ \bibinfo {pages}
  {045108} (\bibinfo {year} {2019})}\BibitemShut {NoStop}%
\bibitem [{\citenamefont {Shen}\ \emph {et~al.}(2020)\citenamefont {Shen},
  \citenamefont {Liu}, \citenamefont {Yu},\ and\ \citenamefont
  {Rubenstein}}]{shen_finite_2020}%
  \BibitemOpen
  \bibfield  {author} {\bibinfo {author} {\bibfnamefont {T.}~\bibnamefont
  {Shen}}, \bibinfo {author} {\bibfnamefont {Y.}~\bibnamefont {Liu}}, \bibinfo
  {author} {\bibfnamefont {Y.}~\bibnamefont {Yu}}, \ and\ \bibinfo {author}
  {\bibfnamefont {B.~M.}\ \bibnamefont {Rubenstein}},\ }\href {\doibase
  10.1063/5.0026606} {\bibfield  {journal} {\bibinfo  {journal} {The Journal of
  Chemical Physics}\ }\textbf {\bibinfo {volume} {153}},\ \bibinfo {pages}
  {204108} (\bibinfo {year} {2020})}\BibitemShut {NoStop}%
\bibitem [{\citenamefont {Church}\ and\ \citenamefont
  {Rubenstein}(2021)}]{church_real-time_2021}%
  \BibitemOpen
  \bibfield  {author} {\bibinfo {author} {\bibfnamefont {M.~S.}\ \bibnamefont
  {Church}}\ and\ \bibinfo {author} {\bibfnamefont {B.~M.}\ \bibnamefont
  {Rubenstein}},\ }\href {\doibase 10.1063/5.0049116} {\bibfield  {journal}
  {\bibinfo  {journal} {The Journal of Chemical Physics}\ }\textbf {\bibinfo
  {volume} {154}},\ \bibinfo {pages} {184103} (\bibinfo {year}
  {2021})}\BibitemShut {NoStop}%
\bibitem [{\citenamefont {Liu}\ \emph {et~al.}(2020)\citenamefont {Liu},
  \citenamefont {Shen}, \citenamefont {Zhang},\ and\ \citenamefont
  {Rubenstein}}]{liu_unveiling_2020}%
  \BibitemOpen
  \bibfield  {author} {\bibinfo {author} {\bibfnamefont {Y.}~\bibnamefont
  {Liu}}, \bibinfo {author} {\bibfnamefont {T.}~\bibnamefont {Shen}}, \bibinfo
  {author} {\bibfnamefont {H.}~\bibnamefont {Zhang}}, \ and\ \bibinfo {author}
  {\bibfnamefont {B.}~\bibnamefont {Rubenstein}},\ }\href {\doibase
  10.1021/acs.jctc.0c00288} {\bibfield  {journal} {\bibinfo  {journal} {Journal
  of Chemical Theory and Computation}\ }\textbf {\bibinfo {volume} {16}},\
  \bibinfo {pages} {4298} (\bibinfo {year} {2020})}\BibitemShut {NoStop}%
\bibitem [{\citenamefont {Blunt}, \citenamefont {Alavi},\ and\ \citenamefont
  {Booth}(2015)}]{blunt_krylov-projected_2015}%
  \BibitemOpen
  \bibfield  {author} {\bibinfo {author} {\bibfnamefont {N.}~\bibnamefont
  {Blunt}}, \bibinfo {author} {\bibfnamefont {A.}~\bibnamefont {Alavi}}, \ and\
  \bibinfo {author} {\bibfnamefont {G.~H.}\ \bibnamefont {Booth}},\ }\href
  {\doibase 10.1103/PhysRevLett.115.050603} {\bibfield  {journal} {\bibinfo
  {journal} {Physical Review Letters}\ }\textbf {\bibinfo {volume} {115}},\
  \bibinfo {pages} {050603} (\bibinfo {year} {2015})}\BibitemShut {NoStop}%
\bibitem [{\citenamefont {Blunt}\ \emph {et~al.}(2014)\citenamefont {Blunt},
  \citenamefont {Rogers}, \citenamefont {Spencer},\ and\ \citenamefont
  {Foulkes}}]{blunt_density-matrix_2014}%
  \BibitemOpen
  \bibfield  {author} {\bibinfo {author} {\bibfnamefont {N.~S.}\ \bibnamefont
  {Blunt}}, \bibinfo {author} {\bibfnamefont {T.~W.}\ \bibnamefont {Rogers}},
  \bibinfo {author} {\bibfnamefont {J.~S.}\ \bibnamefont {Spencer}}, \ and\
  \bibinfo {author} {\bibfnamefont {W.~M.~C.}\ \bibnamefont {Foulkes}},\ }\href
  {\doibase 10.1103/PhysRevB.89.245124} {\bibfield  {journal} {\bibinfo
  {journal} {Physical Review B}\ }\textbf {\bibinfo {volume} {89}},\ \bibinfo
  {pages} {245124} (\bibinfo {year} {2014})}\BibitemShut {NoStop}%
\bibitem [{\citenamefont {Booth}, \citenamefont {Thom},\ and\ \citenamefont
  {Alavi}(2009)}]{booth_fermion_2009}%
  \BibitemOpen
  \bibfield  {author} {\bibinfo {author} {\bibfnamefont {G.~H.}\ \bibnamefont
  {Booth}}, \bibinfo {author} {\bibfnamefont {A.~J.~W.}\ \bibnamefont {Thom}},
  \ and\ \bibinfo {author} {\bibfnamefont {A.}~\bibnamefont {Alavi}},\ }\href
  {\doibase 10.1063/1.3193710} {\bibfield  {journal} {\bibinfo  {journal} {The
  Journal of Chemical Physics}\ }\textbf {\bibinfo {volume} {131}},\ \bibinfo
  {pages} {054106} (\bibinfo {year} {2009})}\BibitemShut {NoStop}%
\bibitem [{\citenamefont {Malone}\ \emph {et~al.}(2015)\citenamefont {Malone},
  \citenamefont {Blunt}, \citenamefont {Shepherd}, \citenamefont {Lee},
  \citenamefont {Spencer},\ and\ \citenamefont
  {Foulkes}}]{malone_interaction_2015}%
  \BibitemOpen
  \bibfield  {author} {\bibinfo {author} {\bibfnamefont {F.~D.}\ \bibnamefont
  {Malone}}, \bibinfo {author} {\bibfnamefont {N.~S.}\ \bibnamefont {Blunt}},
  \bibinfo {author} {\bibfnamefont {J.~J.}\ \bibnamefont {Shepherd}}, \bibinfo
  {author} {\bibfnamefont {D.~K.~K.}\ \bibnamefont {Lee}}, \bibinfo {author}
  {\bibfnamefont {J.~S.}\ \bibnamefont {Spencer}}, \ and\ \bibinfo {author}
  {\bibfnamefont {W.~M.~C.}\ \bibnamefont {Foulkes}},\ }\href {\doibase
  10.1063/1.4927434} {\bibfield  {journal} {\bibinfo  {journal} {The Journal of
  Chemical Physics}\ }\textbf {\bibinfo {volume} {143}},\ \bibinfo {pages}
  {044116} (\bibinfo {year} {2015})}\BibitemShut {NoStop}%
\bibitem [{\citenamefont {Cleland}, \citenamefont {Booth},\ and\ \citenamefont
  {Alavi}(2010)}]{cleland_communications_2010}%
  \BibitemOpen
  \bibfield  {author} {\bibinfo {author} {\bibfnamefont {D.}~\bibnamefont
  {Cleland}}, \bibinfo {author} {\bibfnamefont {G.~H.}\ \bibnamefont {Booth}},
  \ and\ \bibinfo {author} {\bibfnamefont {A.}~\bibnamefont {Alavi}},\ }\href
  {\doibase 10.1063/1.3302277} {\bibfield  {journal} {\bibinfo  {journal} {The
  Journal of Chemical Physics}\ }\textbf {\bibinfo {volume} {132}},\ \bibinfo
  {pages} {041103} (\bibinfo {year} {2010})}\BibitemShut {NoStop}%
\bibitem [{\citenamefont {Dornheim}\ \emph {et~al.}(2017)\citenamefont
  {Dornheim}, \citenamefont {Groth}, \citenamefont {Malone}, \citenamefont
  {Schoof}, \citenamefont {Sjostrom}, \citenamefont {Foulkes},\ and\
  \citenamefont {Bonitz}}]{dornheim_ab_2017}%
  \BibitemOpen
  \bibfield  {author} {\bibinfo {author} {\bibfnamefont {T.}~\bibnamefont
  {Dornheim}}, \bibinfo {author} {\bibfnamefont {S.}~\bibnamefont {Groth}},
  \bibinfo {author} {\bibfnamefont {F.~D.}\ \bibnamefont {Malone}}, \bibinfo
  {author} {\bibfnamefont {T.}~\bibnamefont {Schoof}}, \bibinfo {author}
  {\bibfnamefont {T.}~\bibnamefont {Sjostrom}}, \bibinfo {author}
  {\bibfnamefont {W.~M.~C.}\ \bibnamefont {Foulkes}}, \ and\ \bibinfo {author}
  {\bibfnamefont {M.}~\bibnamefont {Bonitz}},\ }\href {\doibase
  10.1063/1.4977920} {\bibfield  {journal} {\bibinfo  {journal} {Physics of
  Plasmas}\ }\textbf {\bibinfo {volume} {24}},\ \bibinfo {pages} {056303}
  (\bibinfo {year} {2017})}\BibitemShut {NoStop}%
\bibitem [{\citenamefont {Groth}\ \emph {et~al.}(2017)\citenamefont {Groth},
  \citenamefont {Dornheim}, \citenamefont {Sjostrom}, \citenamefont {Malone},
  \citenamefont {Foulkes},\ and\ \citenamefont {Bonitz}}]{groth_ab_2017}%
  \BibitemOpen
  \bibfield  {author} {\bibinfo {author} {\bibfnamefont {S.}~\bibnamefont
  {Groth}}, \bibinfo {author} {\bibfnamefont {T.}~\bibnamefont {Dornheim}},
  \bibinfo {author} {\bibfnamefont {T.}~\bibnamefont {Sjostrom}}, \bibinfo
  {author} {\bibfnamefont {F.~D.}\ \bibnamefont {Malone}}, \bibinfo {author}
  {\bibfnamefont {W.}~\bibnamefont {Foulkes}}, \ and\ \bibinfo {author}
  {\bibfnamefont {M.}~\bibnamefont {Bonitz}},\ }\href {\doibase
  10.1103/PhysRevLett.119.135001} {\bibfield  {journal} {\bibinfo  {journal}
  {Physical Review Letters}\ }\textbf {\bibinfo {volume} {119}},\ \bibinfo
  {pages} {135001} (\bibinfo {year} {2017})}\BibitemShut {NoStop}%
\bibitem [{\citenamefont {Dornheim}\ \emph {et~al.}(2016)\citenamefont
  {Dornheim}, \citenamefont {Groth}, \citenamefont {Sjostrom}, \citenamefont
  {Malone}, \citenamefont {Foulkes},\ and\ \citenamefont
  {Bonitz}}]{dornheim_ab_2016}%
  \BibitemOpen
  \bibfield  {author} {\bibinfo {author} {\bibfnamefont {T.}~\bibnamefont
  {Dornheim}}, \bibinfo {author} {\bibfnamefont {S.}~\bibnamefont {Groth}},
  \bibinfo {author} {\bibfnamefont {T.}~\bibnamefont {Sjostrom}}, \bibinfo
  {author} {\bibfnamefont {F.~D.}\ \bibnamefont {Malone}}, \bibinfo {author}
  {\bibfnamefont {W.}~\bibnamefont {Foulkes}}, \ and\ \bibinfo {author}
  {\bibfnamefont {M.}~\bibnamefont {Bonitz}},\ }\href {\doibase
  10.1103/PhysRevLett.117.156403} {\bibfield  {journal} {\bibinfo  {journal}
  {Physical Review Letters}\ }\textbf {\bibinfo {volume} {117}},\ \bibinfo
  {pages} {156403} (\bibinfo {year} {2016})}\BibitemShut {NoStop}%
\bibitem [{\citenamefont {Petras}\ \emph {et~al.}(2020)\citenamefont {Petras},
  \citenamefont {Ramadugu}, \citenamefont {Malone},\ and\ \citenamefont
  {Shepherd}}]{petras_using_2020}%
  \BibitemOpen
  \bibfield  {author} {\bibinfo {author} {\bibfnamefont {H.~R.}\ \bibnamefont
  {Petras}}, \bibinfo {author} {\bibfnamefont {S.~K.}\ \bibnamefont
  {Ramadugu}}, \bibinfo {author} {\bibfnamefont {F.~D.}\ \bibnamefont
  {Malone}}, \ and\ \bibinfo {author} {\bibfnamefont {J.~J.}\ \bibnamefont
  {Shepherd}},\ }\href {\doibase 10.1021/acs.jctc.9b01080} {\bibfield
  {journal} {\bibinfo  {journal} {Journal of Chemical Theory and Computation}\
  }\textbf {\bibinfo {volume} {16}},\ \bibinfo {pages} {1029} (\bibinfo {year}
  {2020})}\BibitemShut {NoStop}%
\bibitem [{\citenamefont {Petras}\ \emph {et~al.}(2021)\citenamefont {Petras},
  \citenamefont {Van~Benschoten}, \citenamefont {Ramadugu},\ and\ \citenamefont
  {Shepherd}}]{petras_sign_2021}%
  \BibitemOpen
  \bibfield  {author} {\bibinfo {author} {\bibfnamefont {H.~R.}\ \bibnamefont
  {Petras}}, \bibinfo {author} {\bibfnamefont {W.~Z.}\ \bibnamefont
  {Van~Benschoten}}, \bibinfo {author} {\bibfnamefont {S.~K.}\ \bibnamefont
  {Ramadugu}}, \ and\ \bibinfo {author} {\bibfnamefont {J.~J.}\ \bibnamefont
  {Shepherd}},\ }\href {\doibase 10.1021/acs.jctc.1c00078} {\bibfield
  {journal} {\bibinfo  {journal} {Journal of Chemical Theory and Computation}\
  }\textbf {\bibinfo {volume} {17}},\ \bibinfo {pages} {6036} (\bibinfo {year}
  {2021})}\BibitemShut {NoStop}%
\bibitem [{\citenamefont {Chessex}, \citenamefont {Borrelli},\ and\
  \citenamefont {Öttinger}(2022)}]{chessex_fixed_2022}%
  \BibitemOpen
  \bibfield  {author} {\bibinfo {author} {\bibfnamefont {R.}~\bibnamefont
  {Chessex}}, \bibinfo {author} {\bibfnamefont {M.}~\bibnamefont {Borrelli}}, \
  and\ \bibinfo {author} {\bibfnamefont {H.~C.}\ \bibnamefont {Öttinger}},\
  }\href {\doibase 10.48550/ARXIV.2201.01383} {\  (\bibinfo {year} {2022}),\
  10.48550/ARXIV.2201.01383},\ \bibinfo {note} {publisher: arXiv Version
  Number: 1}\BibitemShut {NoStop}%
\bibitem [{\citenamefont {Kou}\ and\ \citenamefont
  {Hirata}(2014)}]{kou_finite-temperature_2014}%
  \BibitemOpen
  \bibfield  {author} {\bibinfo {author} {\bibfnamefont {Z.}~\bibnamefont
  {Kou}}\ and\ \bibinfo {author} {\bibfnamefont {S.}~\bibnamefont {Hirata}},\
  }\href {\doibase 10.1007/s00214-014-1487-4} {\bibfield  {journal} {\bibinfo
  {journal} {Theoretical Chemistry Accounts}\ }\textbf {\bibinfo {volume}
  {133}},\ \bibinfo {pages} {1487} (\bibinfo {year} {2014})}\BibitemShut
  {NoStop}%
\bibitem [{\citenamefont {Malone}\ \emph {et~al.}(2016)\citenamefont {Malone},
  \citenamefont {Blunt}, \citenamefont {Brown}, \citenamefont {Lee},
  \citenamefont {Spencer}, \citenamefont {Foulkes},\ and\ \citenamefont
  {Shepherd}}]{malone_accurate_2016}%
  \BibitemOpen
  \bibfield  {author} {\bibinfo {author} {\bibfnamefont {F.~D.}\ \bibnamefont
  {Malone}}, \bibinfo {author} {\bibfnamefont {N.}~\bibnamefont {Blunt}},
  \bibinfo {author} {\bibfnamefont {E.~W.}\ \bibnamefont {Brown}}, \bibinfo
  {author} {\bibfnamefont {D.}~\bibnamefont {Lee}}, \bibinfo {author}
  {\bibfnamefont {J.}~\bibnamefont {Spencer}}, \bibinfo {author} {\bibfnamefont
  {W.}~\bibnamefont {Foulkes}}, \ and\ \bibinfo {author} {\bibfnamefont
  {J.~J.}\ \bibnamefont {Shepherd}},\ }\href {\doibase
  10.1103/PhysRevLett.117.115701} {\bibfield  {journal} {\bibinfo  {journal}
  {Physical Review Letters}\ }\textbf {\bibinfo {volume} {117}},\ \bibinfo
  {pages} {115701} (\bibinfo {year} {2016})}\BibitemShut {NoStop}%
\bibitem [{\citenamefont {Cleland}, \citenamefont {Booth},\ and\ \citenamefont
  {Alavi}(2011)}]{cleland_study_2011}%
  \BibitemOpen
  \bibfield  {author} {\bibinfo {author} {\bibfnamefont {D.~M.}\ \bibnamefont
  {Cleland}}, \bibinfo {author} {\bibfnamefont {G.~H.}\ \bibnamefont {Booth}},
  \ and\ \bibinfo {author} {\bibfnamefont {A.}~\bibnamefont {Alavi}},\ }\href
  {\doibase 10.1063/1.3525712} {\bibfield  {journal} {\bibinfo  {journal} {The
  Journal of Chemical Physics}\ }\textbf {\bibinfo {volume} {134}},\ \bibinfo
  {pages} {024112} (\bibinfo {year} {2011})}\BibitemShut {NoStop}%
\bibitem [{\citenamefont {Cleland}\ \emph {et~al.}(2012)\citenamefont
  {Cleland}, \citenamefont {Booth}, \citenamefont {Overy},\ and\ \citenamefont
  {Alavi}}]{cleland_taming_2012}%
  \BibitemOpen
  \bibfield  {author} {\bibinfo {author} {\bibfnamefont {D.}~\bibnamefont
  {Cleland}}, \bibinfo {author} {\bibfnamefont {G.~H.}\ \bibnamefont {Booth}},
  \bibinfo {author} {\bibfnamefont {C.}~\bibnamefont {Overy}}, \ and\ \bibinfo
  {author} {\bibfnamefont {A.}~\bibnamefont {Alavi}},\ }\href {\doibase
  10.1021/ct300504f} {\bibfield  {journal} {\bibinfo  {journal} {Journal of
  Chemical Theory and Computation}\ }\textbf {\bibinfo {volume} {8}},\ \bibinfo
  {pages} {4138} (\bibinfo {year} {2012})}\BibitemShut {NoStop}%
\bibitem [{\citenamefont {Shepherd}, \citenamefont {Booth},\ and\ \citenamefont
  {Alavi}(2012)}]{shepherd_investigation_2012}%
  \BibitemOpen
  \bibfield  {author} {\bibinfo {author} {\bibfnamefont {J.~J.}\ \bibnamefont
  {Shepherd}}, \bibinfo {author} {\bibfnamefont {G.~H.}\ \bibnamefont {Booth}},
  \ and\ \bibinfo {author} {\bibfnamefont {A.}~\bibnamefont {Alavi}},\ }\href
  {\doibase 10.1063/1.4720076} {\bibfield  {journal} {\bibinfo  {journal} {The
  Journal of Chemical Physics}\ }\textbf {\bibinfo {volume} {136}},\ \bibinfo
  {pages} {244101} (\bibinfo {year} {2012})}\BibitemShut {NoStop}%
\bibitem [{\citenamefont {Booth}\ \emph {et~al.}(2013)\citenamefont {Booth},
  \citenamefont {Grüneis}, \citenamefont {Kresse},\ and\ \citenamefont
  {Alavi}}]{booth_towards_2013}%
  \BibitemOpen
  \bibfield  {author} {\bibinfo {author} {\bibfnamefont {G.~H.}\ \bibnamefont
  {Booth}}, \bibinfo {author} {\bibfnamefont {A.}~\bibnamefont {Grüneis}},
  \bibinfo {author} {\bibfnamefont {G.}~\bibnamefont {Kresse}}, \ and\ \bibinfo
  {author} {\bibfnamefont {A.}~\bibnamefont {Alavi}},\ }\href {\doibase
  10.1038/nature11770} {\bibfield  {journal} {\bibinfo  {journal} {Nature}\
  }\textbf {\bibinfo {volume} {493}},\ \bibinfo {pages} {365} (\bibinfo {year}
  {2013})}\BibitemShut {NoStop}%
\bibitem [{\citenamefont {Thomas}, \citenamefont {Booth},\ and\ \citenamefont
  {Alavi}(2015)}]{thomas_accurate_2015}%
  \BibitemOpen
  \bibfield  {author} {\bibinfo {author} {\bibfnamefont {R.~E.}\ \bibnamefont
  {Thomas}}, \bibinfo {author} {\bibfnamefont {G.~H.}\ \bibnamefont {Booth}}, \
  and\ \bibinfo {author} {\bibfnamefont {A.}~\bibnamefont {Alavi}},\ }\href
  {\doibase 10.1103/PhysRevLett.114.033001} {\bibfield  {journal} {\bibinfo
  {journal} {Physical Review Letters}\ }\textbf {\bibinfo {volume} {114}},\
  \bibinfo {pages} {033001} (\bibinfo {year} {2015})}\BibitemShut {NoStop}%
\bibitem [{GCI()}]{GCI}%
  \BibitemOpen
  \href@noop {} {}\bibinfo {note} {In the grand canonical initialization scheme
  introduced at the same time as IP-DMQMC\cite{malone_interaction_2015}, the 
  walker number was interpreted as the number of times initialization was
  attempted. For this paper, this was modified to instead count the number
  of walkers created}\BibitemShut {NoStop}%
\bibitem [{\citenamefont {Malone}(2017)}]{malone_quantum_2017}%
  \BibitemOpen
  \bibfield  {author} {\bibinfo {author} {\bibfnamefont {F.~D.}\ \bibnamefont
  {Malone}},\ }\emph {\bibinfo {title} {Quantum {Monte} {Carlo} {Simulations}
  of {Warm} {Dense} {Matter}}},\ \href@noop {} {Ph.D. thesis} (\bibinfo {year}
  {2017})\BibitemShut {NoStop}%
\bibitem [{\citenamefont {Spencer}\ \emph {et~al.}(2019)\citenamefont
  {Spencer}, \citenamefont {Blunt}, \citenamefont {Choi}, \citenamefont
  {Etrych}, \citenamefont {Filip}, \citenamefont {Foulkes}, \citenamefont
  {Franklin}, \citenamefont {Handley}, \citenamefont {Malone}, \citenamefont
  {Neufeld}, \citenamefont {Di~Remigio}, \citenamefont {Rogers}, \citenamefont
  {Scott}, \citenamefont {Shepherd}, \citenamefont {Vigor}, \citenamefont
  {Weston}, \citenamefont {Xu},\ and\ \citenamefont
  {Thom}}]{spencer_hande-qmc_2019}%
  \BibitemOpen
  \bibfield  {author} {\bibinfo {author} {\bibfnamefont {J.~S.}\ \bibnamefont
  {Spencer}}, \bibinfo {author} {\bibfnamefont {N.~S.}\ \bibnamefont {Blunt}},
  \bibinfo {author} {\bibfnamefont {S.}~\bibnamefont {Choi}}, \bibinfo {author}
  {\bibfnamefont {J.}~\bibnamefont {Etrych}}, \bibinfo {author} {\bibfnamefont
  {M.-A.}\ \bibnamefont {Filip}}, \bibinfo {author} {\bibfnamefont {W.~M.~C.}\
  \bibnamefont {Foulkes}}, \bibinfo {author} {\bibfnamefont {R.~S.~T.}\
  \bibnamefont {Franklin}}, \bibinfo {author} {\bibfnamefont {W.~J.}\
  \bibnamefont {Handley}}, \bibinfo {author} {\bibfnamefont {F.~D.}\
  \bibnamefont {Malone}}, \bibinfo {author} {\bibfnamefont {V.~A.}\
  \bibnamefont {Neufeld}}, \bibinfo {author} {\bibfnamefont {R.}~\bibnamefont
  {Di~Remigio}}, \bibinfo {author} {\bibfnamefont {T.~W.}\ \bibnamefont
  {Rogers}}, \bibinfo {author} {\bibfnamefont {C.~J.~C.}\ \bibnamefont
  {Scott}}, \bibinfo {author} {\bibfnamefont {J.~J.}\ \bibnamefont {Shepherd}},
  \bibinfo {author} {\bibfnamefont {W.~A.}\ \bibnamefont {Vigor}}, \bibinfo
  {author} {\bibfnamefont {J.}~\bibnamefont {Weston}}, \bibinfo {author}
  {\bibfnamefont {R.}~\bibnamefont {Xu}}, \ and\ \bibinfo {author}
  {\bibfnamefont {A.~J.~W.}\ \bibnamefont {Thom}},\ }\href {\doibase
  10.1021/acs.jctc.8b01217} {\bibfield  {journal} {\bibinfo  {journal} {Journal
  of Chemical Theory and Computation}\ }\textbf {\bibinfo {volume} {15}},\
  \bibinfo {pages} {1728} (\bibinfo {year} {2019})}\BibitemShut {NoStop}%
\bibitem [{\citenamefont {Bernath}\ \emph {et~al.}(2002)\citenamefont
  {Bernath}, \citenamefont {Shayesteh}, \citenamefont {Tereszchuk},\ and\
  \citenamefont {Colin}}]{bernath_vibration-rotation_2002}%
  \BibitemOpen
  \bibfield  {author} {\bibinfo {author} {\bibfnamefont {P.~F.}\ \bibnamefont
  {Bernath}}, \bibinfo {author} {\bibfnamefont {A.}~\bibnamefont {Shayesteh}},
  \bibinfo {author} {\bibfnamefont {K.}~\bibnamefont {Tereszchuk}}, \ and\
  \bibinfo {author} {\bibfnamefont {R.}~\bibnamefont {Colin}},\ }\href
  {\doibase 10.1126/science.1074580} {\bibfield  {journal} {\bibinfo  {journal}
  {Science}\ }\textbf {\bibinfo {volume} {297}},\ \bibinfo {pages} {1323}
  (\bibinfo {year} {2002})}\BibitemShut {NoStop}%
\bibitem [{\citenamefont {Herzberg}(1966)}]{herzberg_electronic_1966}%
  \BibitemOpen
  \bibfield  {author} {\bibinfo {author} {\bibfnamefont {G.}~\bibnamefont
  {Herzberg}},\ }\href@noop {} {\emph {\bibinfo {title} {Electronic spectra and
  electronic structure of polyatomic molecules}}},\ Vol.~\bibinfo {volume} {3}\
  (\bibinfo  {publisher} {van Nostrand},\ \bibinfo {year} {1966})\BibitemShut
  {NoStop}%
\bibitem [{\citenamefont {Frisch}\ \emph {et~al.}(2009)\citenamefont {Frisch},
  \citenamefont {Trucks}, \citenamefont {Schlegel}, \citenamefont {Scuseria},
  \citenamefont {Robb}, \citenamefont {Cheeseman}, \citenamefont {Scalmani},
  \citenamefont {Barone}, \citenamefont {Mennucci}, \citenamefont {Petersson},\
  and\ \citenamefont {{others}}}]{frisch_gaussian_2009}%
  \BibitemOpen
  \bibfield  {author} {\bibinfo {author} {\bibfnamefont {M.}~\bibnamefont
  {Frisch}}, \bibinfo {author} {\bibfnamefont {G.}~\bibnamefont {Trucks}},
  \bibinfo {author} {\bibfnamefont {H.~B.}\ \bibnamefont {Schlegel}}, \bibinfo
  {author} {\bibfnamefont {G.~E.}\ \bibnamefont {Scuseria}}, \bibinfo {author}
  {\bibfnamefont {M.~A.}\ \bibnamefont {Robb}}, \bibinfo {author}
  {\bibfnamefont {J.~R.}\ \bibnamefont {Cheeseman}}, \bibinfo {author}
  {\bibfnamefont {G.}~\bibnamefont {Scalmani}}, \bibinfo {author}
  {\bibfnamefont {V.}~\bibnamefont {Barone}}, \bibinfo {author} {\bibfnamefont
  {B.}~\bibnamefont {Mennucci}}, \bibinfo {author} {\bibfnamefont
  {G.}~\bibnamefont {Petersson}}, \ and\ \bibinfo {author} {\bibnamefont
  {{others}}},\ }\href@noop {} {\bibfield  {journal} {\bibinfo  {journal}
  {Inc., Wallingford CT}\ }\textbf {\bibinfo {volume} {201}} (\bibinfo {year}
  {2009})}\BibitemShut {NoStop}%
\bibitem [{\citenamefont {Lovas}(2002)}]{lovas_diatomic_2002}%
  \BibitemOpen
  \bibfield  {author} {\bibinfo {author} {\bibfnamefont {F.}~\bibnamefont
  {Lovas}},\ }\href {\doibase 10.18434/T4T59X} {\enquote {\bibinfo {title}
  {Diatomic {Spectral} {Database}, {NIST} {Standard} {Reference} {Database}
  114},}\ } (\bibinfo {year} {2002}),\ \bibinfo {note} {type:
  dataset}\BibitemShut {NoStop}%
\bibitem [{\citenamefont {Wharton}\ \emph {et~al.}(1963)\citenamefont
  {Wharton}, \citenamefont {Klemperer}, \citenamefont {Gold}, \citenamefont
  {Strauch}, \citenamefont {Gallagher},\ and\ \citenamefont
  {Derr}}]{wharton_microwave_1963}%
  \BibitemOpen
  \bibfield  {author} {\bibinfo {author} {\bibfnamefont {L.}~\bibnamefont
  {Wharton}}, \bibinfo {author} {\bibfnamefont {W.}~\bibnamefont {Klemperer}},
  \bibinfo {author} {\bibfnamefont {L.~P.}\ \bibnamefont {Gold}}, \bibinfo
  {author} {\bibfnamefont {R.}~\bibnamefont {Strauch}}, \bibinfo {author}
  {\bibfnamefont {J.~J.}\ \bibnamefont {Gallagher}}, \ and\ \bibinfo {author}
  {\bibfnamefont {V.~E.}\ \bibnamefont {Derr}},\ }\href {\doibase
  10.1063/1.1733824} {\bibfield  {journal} {\bibinfo  {journal} {The Journal of
  Chemical Physics}\ }\textbf {\bibinfo {volume} {38}},\ \bibinfo {pages}
  {1203} (\bibinfo {year} {1963})}\BibitemShut {NoStop}%
\bibitem [{\citenamefont {Johnson}\ and\ \citenamefont
  {{others}}(2006)}]{johnson_nist_2006}%
  \BibitemOpen
  \bibfield  {author} {\bibinfo {author} {\bibfnamefont {R.~D.}\ \bibnamefont
  {Johnson}}\ and\ \bibinfo {author} {\bibnamefont {{others}}},\ }\href@noop {}
  {\bibfield  {journal} {\bibinfo  {journal} {http://srdata.nist.gov/cccbdb}\ }
  (\bibinfo {year} {2006})}\BibitemShut {NoStop}%
\bibitem [{\citenamefont {Werner}\ \emph {et~al.}(2019)\citenamefont {Werner},
  \citenamefont {Knowles}, \citenamefont {Knizia}, \citenamefont {Manby},
  \citenamefont {Schütz},\ and\ \citenamefont
  {{others}}}]{werner_molpro_2019}%
  \BibitemOpen
  \bibfield  {author} {\bibinfo {author} {\bibfnamefont {H.-J.}\ \bibnamefont
  {Werner}}, \bibinfo {author} {\bibfnamefont {P.~J.}\ \bibnamefont {Knowles}},
  \bibinfo {author} {\bibfnamefont {G.}~\bibnamefont {Knizia}}, \bibinfo
  {author} {\bibfnamefont {F.~R.}\ \bibnamefont {Manby}}, \bibinfo {author}
  {\bibfnamefont {M.}~\bibnamefont {Schütz}}, \ and\ \bibinfo {author}
  {\bibnamefont {{others}}},\ }\href@noop {} {\enquote {\bibinfo {title}
  {{MOLPRO}, 2019.2 , a package of ab initio programs},}\ } (\bibinfo {year}
  {2019}),\ \bibinfo {note} {see https://www.molpro.net}\BibitemShut {NoStop}%
\bibitem [{\citenamefont {Booth}\ \emph {et~al.}(2011)\citenamefont {Booth},
  \citenamefont {Cleland}, \citenamefont {Thom},\ and\ \citenamefont
  {Alavi}}]{booth_breaking_2011}%
  \BibitemOpen
  \bibfield  {author} {\bibinfo {author} {\bibfnamefont {G.~H.}\ \bibnamefont
  {Booth}}, \bibinfo {author} {\bibfnamefont {D.}~\bibnamefont {Cleland}},
  \bibinfo {author} {\bibfnamefont {A.~J.~W.}\ \bibnamefont {Thom}}, \ and\
  \bibinfo {author} {\bibfnamefont {A.}~\bibnamefont {Alavi}},\ }\href
  {\doibase 10.1063/1.3624383} {\bibfield  {journal} {\bibinfo  {journal} {The
  Journal of Chemical Physics}\ }\textbf {\bibinfo {volume} {135}},\ \bibinfo
  {pages} {084104} (\bibinfo {year} {2011})}\BibitemShut {NoStop}%
\bibitem [{\citenamefont {Van~Benschoten}\ and\ \citenamefont
  {Shepherd}()}]{vanbenschoten_row_2022}%
  \BibitemOpen
  \bibfield  {author} {\bibinfo {author} {\bibfnamefont {W.~Z.}\ \bibnamefont
  {Van~Benschoten}}\ and\ \bibinfo {author} {\bibfnamefont {J.~J.}\
  \bibnamefont {Shepherd}},\ }\href@noop {} {}\bibinfo {note}
  {Unpublished}\BibitemShut {NoStop}%
\end{thebibliography}
 \end{document}